\documentclass[10pt,a4paper,twoside=semi,twocolumn,mpinclude=false,footinclude=false]{scrartcl}
\usepackage[left=1.5cm,right=2.2cm,top=3.4cm,bottom=3cm]{geometry}
\usepackage[mono=false]{libertine}
\usepackage{inconsolata}
\usepackage{roboto}
\usepackage[utf8]{inputenc}
\usepackage{amstext}
\usepackage{amsmath}
\usepackage{bbm}
\usepackage{theorem}
\usepackage{graphicx}
\usepackage{tikz}
\usepackage{pgfbaseplot}
\usepackage[backend=biber,style=numeric,sorting=nyt,maxnames=4]{biblatex}
\usepackage{fancyhdr}
\usepackage[noend]{algpseudocode}
\usepackage{xparse}
\usepackage{xcolor}
\usepackage{colorschemes}
\usepackage{xspace}
\usepackage{microtype}
\usepackage{booktabs}
\usepackage[many]{tcolorbox}
\usepackage{varwidth}
\usepackage[final,color,notref,notcite]{showkeys}
\usepackage[switch]{lineno}
\usepackage{tocbasic}

\DeclareFieldFormat
  [article,inbook,incollection,inproceedings,patent,thesis,unpublished,misc]
  {title}{\mkbibquote{{\textit{#1}}}}

\definecolor{MIScol2}{HTML}{006373}
\colorlet{algotitlebg}{Aluminium3}
\colorlet{algotitlefg}{black}
\colorlet{algocontentbg}{Aluminium1!10!white}
\colorlet{algocontentfg}{black}
\colorlet{algoframe}{Aluminium3}
\colorlet{spotcolor}{MIScol2!66!black}
\newcommand{\spotcolor}{\color{spotcolor}}

\addtokomafont{title}{\sffamily\spotcolor}
\addtokomafont{paragraph}{\spotcolor}
\addtokomafont{section}{\spotcolor}
\addtokomafont{subsection}{\spotcolor}
\addtokomafont{subsubsection}{\spotcolor}
\addtokomafont{descriptionlabel}{\spotcolor}
\setkomafont{caption}{}
\setkomafont{captionlabel}{}
\KOMAoptions{numbers=noenddot}

\colorlet{refkey}{Aluminium4}
\colorlet{labelkey}{ScarletRed3}

\renewcommand{\epsilon}{\varepsilon}
\renewcommand{\phi}{\varphi}

\NewDocumentCommand{\mc}{m}{\ensuremath{\mathcal{#1}}\xspace}
\NewDocumentCommand{\mcH}{}{\ensuremath{\mc{H}}\xspace}
\NewDocumentCommand{\mcHbold}{}{\ensuremath{\mc{H}\kern-0.825em\mc{H}}\xspace}
\NewDocumentCommand{\mcUH}{}{\ensuremath{\mc{U}\mc{H}}\xspace}
\NewDocumentCommand{\mcHH}{}{\ensuremath{\mc{H}^2}\xspace}
\NewDocumentCommand{\mcO}{}{\ensuremath{\mc{O}}\xspace}

\NewDocumentCommand{\mcL}{}{\ensuremath{\mc{L}}\xspace}
\NewDocumentCommand{\mcS}{}{\ensuremath{\mc{S}}\xspace}

\NewDocumentCommand{\mcW}{}{\ensuremath{\mc{W}}\xspace}
\NewDocumentCommand{\mcX}{}{\ensuremath{\mc{X}}\xspace}

\NewDocumentCommand{\mcM}{}{\ensuremath{\mc{M}}\xspace}

\NewDocumentCommand{\landau}{m}{\ensuremath{\mcO\left(#1\right)}\xspace}
\NewDocumentCommand{\set}{m}{\ensuremath{\left\{ #1 \right\}}\xspace}

\NewDocumentCommand{\rank}{}{\ensuremath{\operatorname{rank}}\xspace}
\NewDocumentCommand{\level}{}{\ensuremath{\operatorname{level}}\xspace}

\NewDocumentCommand{\N}{}{\ensuremath{\mathbbm{N}}}
\NewDocumentCommand{\R}{}{\ensuremath{\mathbbm{R}}}

\NewDocumentCommand{\op}{m}{\ensuremath{\operatorname{#1}}\xspace}

\NewDocumentCommand{\troot}{m}{\ensuremath{\op{root}(#1)}\xspace}
\NewDocumentCommand{\sons}{m}{\ensuremath{\op{sons}(#1)}\xspace}

\NewDocumentCommand{\function}{ m o }{\textnormal{{\ttfamily\textsc{#1}\IfNoValueF{#2}{(#2)}}}\xspace}

\newtcolorbox[auto counter]{algorithmbox}[2][]{  
  floatplacement=htb,
  float,
  enhanced,
  size=fbox,
  before skip=3mm,
  colbacktitle=algotitlebg!25!white,
  title style={top color=algotitlebg!20!white,bottom color=algotitlebg!30!white},
  colback=algocontentbg,
  colframe=algoframe,
  arc=1pt,
  halign=flush left,
  coltitle=algotitlefg,
  title={\textbf{Algorithm~\thetcbcounter:} #2},
  label={#1}}
\newtcolorbox[auto counter]{inlinealgorithmbox}{ 
  size=tight,
  boxsep=0.5mm,
  before skip=2mm,
  after skip=2mm,
  colback=algocontentbg,
  colframe=algoframe,
  arc=1pt,
  halign=flush left}
\newtcolorbox[auto counter]{figurealgorithmbox}{ 
  size=tight,
  before skip=0mm,
  colback=algocontentbg,
  boxrule=0pt,
  colframe=white,
  frame hidden,
  halign=flush left}
\NewDocumentEnvironment{inlinealgorithm}%
                       {}%
                       {\begin{inlinealgorithmbox}
                           \hspace{-0.9em}\begin{varwidth}{\linewidth}\ttfamily\small%
                           \begin{algorithmic}}%
                       {\end{algorithmic}\end{varwidth}\end{inlinealgorithmbox}}

\DeclareNewTOC[
  type = fltalgorithm ,
  float ,
]{fltalgorithm}
                 
\NewDocumentEnvironment{algorithm}%
                       {m m}%
                       {\begin{algorithmbox}[#2]{#1}%
                           \hspace{-1em}\begin{varwidth}{\linewidth}\ttfamily\small%
                           \begin{algorithmic}}%
                       {\end{algorithmic}\end{varwidth}\end{algorithmbox}}
\NewDocumentEnvironment{algorithm*}%
                       {m m}%
                       {\begin{fltalgorithm*}\begin{algorithmbox}[#2]{#1}%
                           \hspace{-1em}\begin{varwidth}{\linewidth}\ttfamily\small%
                           \begin{algorithmic}}%
                       {\end{algorithmic}\end{varwidth}\end{algorithmbox}\end{fltalgorithm*}}

\NewDocumentEnvironment{figurealgorithm}%
                       {}%
                       {\begin{figurealgorithmbox}
                           \hspace{-.75em}\begin{varwidth}{\linewidth}\ttfamily\small%
                           \begin{algorithmic}}%
                       {\end{algorithmic}\end{varwidth}\end{figurealgorithmbox}}

\algnewcommand{\SFor}[2]{\State \algorithmicfor\ #1\ \algorithmicdo\ #2\ }
\algnewcommand{\SForAll}[2]{\State \algorithmicforall\ #1\ \algorithmicdo\ #2\ }
\algnewcommand{\SIf}[2]{\State \algorithmicif\ #1\ \algorithmicthen\ #2\ }
\algnewcommand{\SElsIf}[2]{\State \algorithmicelse\ \algorithmicif\ #1\ \algorithmicthen\ #2\ }
\algnewcommand{\SElse}[1]{\State \algorithmicelse\ #1\ }

\colorlet{colinadm}{ScarletRed2!33!white}
\colorlet{coladm}{Chameleon3}

\NewDocumentCommand{\drawlr}{}{%
  \fill [coladm,opacity=0.5] (0,0) rectangle ++(1cm,0.25cm);
  \fill [coladm,opacity=0.5] (0,0) rectangle ++(0.25cm,1cm);
}

\NewDocumentCommand{\drawcoupling}{}{%
  \fill [colinadm] (0,0) rectangle ++(0.25cm,0.25cm);
}

\NewDocumentCommand{\drawhtree}{}{%
  \begin{scope}[yscale=-1]
    \foreach \ofs in {0cm,0.5cm,1cm,1.5cm,...,3cm} {
      \begin{scope}[xshift=\ofs,yshift=\ofs,scale=0.25]
        \fill[colinadm] (0,0) rectangle ++(2,2);
        \fill[colinadm] (1,1) rectangle ++(2,2);
        \fill[colinadm] (2,2) rectangle ++(2,2);
        \draw[Aluminium4,line width=0.25pt] (0,0) grid ++(4,4);
      \end{scope}
    }

    \foreach \ofs in {0cm,1cm,2cm} {
      \begin{scope}[xshift=\ofs,yshift=\ofs,scale=0.5]
        \draw[Aluminium5,line width=0.5pt] (0,0) grid ++(4,4);
      \end{scope}
    }
    
    \draw[Aluminium6,line width=0.75pt] (0,0) grid ++(4,4);
  \end{scope}
}

\NewDocumentCommand{\drawhmat}{}{%
  \begin{scope}[yscale=-1]
    \foreach \xofs/\yofs in {2cm/0cm,3cm/0cm,3cm/1cm,0cm/2cm,0cm/3cm,1cm/3cm} {
      \begin{scope}[xshift=\xofs,yshift=\yofs] \drawlr \end{scope}
    }
    \foreach \ofs in {0cm,0.5cm,...,2.5cm} {
      \begin{scope}[xshift=\ofs,yshift=\ofs+1cm,scale=0.5] \drawlr \end{scope}
      \begin{scope}[xshift=\ofs+1cm,yshift=\ofs,scale=0.5] \drawlr \end{scope}
    }
    \foreach \ofs in {0cm,1cm,2cm} {
      \begin{scope}[xshift=\ofs,yshift=\ofs+1.5cm,scale=0.5] \drawlr \end{scope}
      \begin{scope}[xshift=\ofs+1.5cm,yshift=\ofs,scale=0.5] \drawlr \end{scope}
    }
    \foreach \ofs in {0cm,0.25cm,...,3.25cm} {
      \begin{scope}[xshift=\ofs,yshift=\ofs+0.5cm,scale=0.25] \drawlr \end{scope}
      \begin{scope}[xshift=\ofs+0.5cm,yshift=\ofs,scale=0.25] \drawlr \end{scope}
    }
    \foreach \ofs in {0cm,0.5cm,...,3cm} {
      \begin{scope}[xshift=\ofs,yshift=\ofs+0.75cm,scale=0.25] \drawlr \end{scope}
      \begin{scope}[xshift=\ofs+0.75cm,yshift=\ofs,scale=0.25] \drawlr \end{scope}
    }
  \end{scope}
  \drawhtree
}

\NewDocumentCommand{\drawhhmat}{}{%
  \begin{scope}[yscale=-1]
    \foreach \xofs/\yofs in {2cm/0cm,3cm/0cm,3cm/1cm,0cm/2cm,0cm/3cm,1cm/3cm} {
      \begin{scope}[xshift=\xofs,yshift=\yofs] \drawcoupling \end{scope}
    }
    \foreach \ofs in {0cm,0.5cm,...,2.5cm} {
      \begin{scope}[xshift=\ofs,yshift=\ofs+1cm,scale=0.5] \drawcoupling \end{scope}
      \begin{scope}[xshift=\ofs+1cm,yshift=\ofs,scale=0.5] \drawcoupling \end{scope}
    }
    \foreach \ofs in {0cm,1cm,2cm} {
      \begin{scope}[xshift=\ofs,yshift=\ofs+1.5cm,scale=0.5] \drawcoupling \end{scope}
      \begin{scope}[xshift=\ofs+1.5cm,yshift=\ofs,scale=0.5] \drawcoupling \end{scope}
    }
    \foreach \ofs in {0cm,0.25cm,...,3.25cm} {
      \begin{scope}[xshift=\ofs,yshift=\ofs+0.5cm,scale=0.25] \drawcoupling \end{scope}
      \begin{scope}[xshift=\ofs+0.5cm,yshift=\ofs,scale=0.25] \drawcoupling \end{scope}
    }
    \foreach \ofs in {0cm,0.5cm,...,3cm} {
      \begin{scope}[xshift=\ofs,yshift=\ofs+0.75cm,scale=0.25] \drawcoupling \end{scope}
      \begin{scope}[xshift=\ofs+0.75cm,yshift=\ofs,scale=0.25] \drawcoupling \end{scope}
    }
  \end{scope}
  \drawhtree
}

\theoremstyle{plain}
\theoremheaderfont{\bfseries \upshape}
\newtheorem{define}{Definition}[section]
\newtheorem{remark}[define]{Remark}

\setkomafont{author}{\normalsize}
\setkomafont{date}{\normalsize}

\makeatletter
\renewcommand{\maketitle}{
  \twocolumn[{%
  \centering
  \begin{minipage}[t]{.9\linewidth}%
  \centering
  {\usekomafont{title}{\huge\@title\par}}
  \vskip 1.0em
  {\usekomafont{title}{\large\@subtitle\par}}
  \vskip 1.0em
  {\spotcolor \hrule height 2pt}
  \vskip 1.0em
  \usekomafont{author}{%
    \lineskip 0.75em
    \begin{tabular}[t]{c}
      \@author
    \end{tabular}\par
  }%
  \vskip 1.5em
  {\usekomafont{date}{\@date\par}}
  \vskip 1.0em
  {\spotcolor \hrule height 2pt}
  \vskip 1.0em
  \end{minipage}
  \vskip 3.0em
  }]
}
\makeatother

\renewenvironment{abstract}{%
  \centering
  \setlength{\parindent}{0pt}
  \begin{minipage}{\linewidth}%
    \setlength{\parindent}{0pt}%
    \setlength{\parskip}{.5em}%
    \textbf{\textcolor{spotcolor}{Abstract}}%
  }{\end{minipage}
}

\NewDocumentCommand{\clt}{}{\ensuremath{\tau}\xspace}
\NewDocumentCommand{\cls}{}{\ensuremath{\sigma}\xspace}

\title{Floating Point Compression of Hierarchical Matrix Formats and its Impact on Matrix-Vector Multiplication}
\author{{\large Ronald Kriemann}\small\\
  MPI for Mathematics i.t.S.\\
  Leipzig, Germany\\
  rok@mis.mpg.de
}

\begin{document}

\maketitle

\begin{abstract}
  Matrix-vector multiplication forms the basis of many iterative solution algorithms and as such is an
  important algorithm also for hierarchical matrices which are used to represent dense data in an optimized form by
  applying low-rank compression. However, due to its low computational intensity, the performance of matrix-vector
  multiplication is typically limited by the available memory bandwidth on parallel systems. With floating point
  compression the memory footprint can be optimized, which reduces the stress on the memory sub system and thereby
  increases performance. We will look into the compression of different formats of hierachical matrices and how this can
  be used to speed up the corresponding matrix-vector multiplication.
  

\end{abstract}

\pagestyle{fancy}
\thispagestyle{plain}
\fancyhf{}
\fancyhf[HLE]{\footnotesize{\thepage\hfill R. Kriemann}}
\fancyhf[HRO]{\footnotesize{Floating Point Compression of \mcH-Matrix Formats and its Impact on MVM\hfill\thepage}}
\fancyhf[FC]{}
\renewcommand{\headrulewidth}{0.4pt}
\renewcommand{\footrulewidth}{0pt}
\thispagestyle{empty}




\section{Introduction} \label{sec:intro}

As opposed to sparse matrices, dense matrices have a serious storage problem as the amount of required memory scales
quadratically with the number of rows and columns. This translates into an even higher complexity for arithmetic
operations. Therefore, certain structural properties of such matrices have been exploited from the very beginning to
reduces the memory costs. Low-rank techniques, especially in the form of hierarchical matrices (\mcH-matrices,
\cite{Hackbusch:1999}), have demonstrated a high efficiency for this in bringing down the storage complexity to
(almost) linear levels which furthermore also applies to their arithmetic.

While the linear storage (and compute) complexity is the major factor in reduced storage costs, further optimization is
possible for the actual storage format used for representing low-rank data in memory. Here, the double (FP64) or single
(FP32) precision format is used. However, as low-rank approximation and arithmetic is normally associated with a user
defined accuracy, this accuracy is typically much coarser than the unit roundoff of the utilised floating point format
shown in Table~\ref{tab:ieee754}. Therefore, a floating point representation with a precision closer to the
low-rank accuracy is wanted to further reduce storage (and possibly computation) costs.

\begin{table}[tbp]
  \centering
  \begin{tabular}{ll}
    \multicolumn{2}{r}{\textbf{Unit Roundoff}} 
    \\
    \toprule
    FP64 & \(1.11 \times 10^{-16}\) \\
    FP32 & \(5.96 \times 10^{-8}\)  \\
    TF32 & \(4.88 \times 10^{-4}\) \\
    BF16 & \(3.91 \times 10^{-3}\) \\
    FP16 & \(4.88 \times 10^{-4}\) \\
    FP8\footnotemark[1] & \(6.25 \times 10^{-2}\)
  \end{tabular}
  \caption{Unit roundoff for floating point formats of the IEEE-754 standard and related formats.}
  \label{tab:ieee754}
\end{table}

\footnotetext[1]{Using the E4M3 version of the FP8 format.}
Usually, e.g., \cite{AbdCaoPeiBos:2022,CarCheLiu:2024}, this is implemented with \emph{mixed precision} formats, i.e.,
with hardware supported floating point formats like FP16 or BF16 in addition to FP32 and FP64. A major disadvantage if
these formats is the big difference in memory sizes, e.g., two byte for FP16/BF16, four byte for FP32 and eight byte for
FP64, which translates into significant precision gaps, e.g., from about \(10^{-3}\) to \(10^{-8}\) to \(10^{-16}\),
which reduces the flexibility in adjusting to the low-rank accuracy.

Hardware provided floating point formats can be used for a special handling of low-rank matrices where parts of the data
are stored in different formats \cite{OoiIawFuk:2020,AbdCaoPeiBos:2022}. This allows to further reduce memory costs and
thereby also increase performance. Nevertheless, also these approaches are limited by the big jumps in memory sizes
between such hardware supported floating point formats.

Therefore, the scheme in \cite{AbdCaoPeiBos:2022} was combined in \cite{Kri:2025} with general floating point
compression methods which allow a much more fine-grained accuracy setting and hence control of memory sizes. Especially
for low-rank data this leads to a very efficient storage scheme.

Aside from \mcH-matrices other hierarchical matrix formats exist like uniform \mcH-matrices
\cite{Hackbusch:1999,BruHuyMee:2025} or \mcHH-matrices \cite{HacKhoSau:2000,Boerm:2010}, which use a more advanced
low-rank storage scheme with shared or nested bases, requiring only small coefficient matrices for each matrix
block. The compression schemes from \cite{Kri:2025} can also be applied to these matrix formats. The results of this
will be investigated in this work.

Standard \mcH-arithmetic is typically based on dense arithmetic functions defined by the BLAS and LAPACK function
set \cite{lapack}. Therefore, the modified \mcH-arithmetic in \cite{Kri:2025} was based on the idea of decompressing all
input data of arithmetic kernel functions, executing the arithmetic kernel in standard double precision and then
compressing the output data. This way, the actual arithmetic functions remain unchanged. Already this approach showed
superior performance for the \mcH-matrix-vector multiplication (\mcH-MVM), which is often memory bandwidth limited and
as such, any reduction of the memory size will increase performance.

Another reason for this \emph{semi-on-the-fly} approach was the general compression approach in \cite{Kri:2025}, i.e.,
for floating point data \emph{any} compressor could be used, making it more difficult to optimize access to compressed
data.

However, some of the compression schemes in \cite{Kri:2025} allow random access of entries in the compressed storage and
hence, special arithmetic functions can be implemented more easily. The potential benefit of such an approach for
\mcH-matrix vector multiplication for different hierarchical matrix formats will be investigated in this work.

An analogue strategy was used in \cite{AnzGruQui:2019} with the idea of a \emph{memory accessor}, i.e., transparent
conversion between a storage and a computation format within a sparse matrix computation. This work is therefore an
application of this concept for \mcH-matrix arithmetic.

For a special floating point compression scheme this is also investigated in \cite{AmeJegExcMarPic:2025} for dense and
low-rank matrix-vector multiplication with a special focus on performance. We will use an almost identical floating
point format also in this work.

The rest of this work is structured as follows: in Section~\ref{sec:hmat} basic definitions and algorithms for
\mcH-matrices in different formats are introduced. Section~\ref{sec:mvm} will discuss different strategies for
\mcH-matrix-vector multiplication. Compression schemes and their application to data within hierarchical matrices and
their effect on matrix-vector multiplication will be the topic of Section~\ref{sec:zhmat}, followed by a conclusion in
Section~\ref{sec:conclude}.


\section{Hierarchical Matrices} \label{sec:hmat}

\subsection{Model Problem} \label{sec:model}

Our model problem is based on a boundary element discretization for the Laplace single layer potential (Laplace SLP).

Let \(\Gamma = \set{ x \in \R^3 : \|x\|_2 = 1}\) be the domain and \(f: \Gamma \rightarrow \R\) a given right-hand
side. Then the unknown \(u\) is defined by 
\begin{equation} \label{eqn:slp}
  \int_{\Gamma} \frac{1}{\|x-y\|} u(x) dy = f(x), \quad x \in \Gamma
\end{equation}
Using a triangulation \(\Gamma = \cup_{i \in I} \pi_i, I := \set{0,\ldots,n-1}\), the function \(u\) is replaced by a
discrete version \(u_h := \sum_{i \in I} u_i \phi_i\) with (basis) functions
\begin{displaymath}
  \phi_i(x) := \left\{ \begin{matrix} 1 & x \in \pi_i \\ 0 & \textnormal{otherwise} \end{matrix} \right.,
\end{displaymath}
e.g., piecewise constant ansatz functions over \(\Gamma\). Using the Galerkin method this results in an equation system
with a matrix \(M \in \R^{I \times I}\) defined by
\begin{equation} \label{eqn:matcoeff}
  m_{ij} := \int_{\pi_i} \int_{\pi_j} \frac{1}{\|x-y\|} dx dy
\end{equation}
The double integral \eqref{eqn:matcoeff} is approximated using quadrature rules described in \cite[Chapter 9]{SauSch:2011}.
As \eqref{eqn:matcoeff} is non-zero everywhere, \(M\) is a dense matrix which will be approximated by low-rank methods.


\subsection{\mcH-Matrices}

For the indexset \(I\) we define the \emph{cluster tree} as the hierarchical partitioning of \(I\) into
disjoint sub-sets of \(I\):
\begin{define}[Cluster Tree]
  Let \(T_I = (V,E)\) be a tree with \(V \subset \mathcal{P}(I)\). \(T_I\) is called a \emph{cluster tree} over \(I\) if
  \begin{enumerate}
  \item \(I = \troot{T_I}\) and
  \item for all \(v \in V\) with \(\sons{v} \ne \emptyset : v = \dot\cup_{v' \in \sons{v}} v'\).
  \end{enumerate}
  A node in \(T_I\) is also called a \emph{cluster} and we write \(\clt \in T_I\) if \(\clt \in V\). The set of leaves
  of \(T_I\) is denoted by \(\mcL(T_I)\). Furthermore \(\level(\clt)\) denotes the (graph) distance of \(\clt\) from the
  root \(I\).
\end{define}

Similar to a cluster tree we can extend the hierarchical partitioning to the product \(I \times J\) of two index sets
\(I, J\), while restricting the possible set of nodes by given cluster trees \(T_I\) and \(T_J\) over \(I\) and \(J\),
respectively. Furthermore, the set of leaves will be defined by an \emph{admissibility condition}, which is used to
detect blocks in the product tree which can be efficiently approximated by low-rank matrices with a predefined rank
\(k\). In the literature, various examples of admissibility can found, e.g. standard \cite{HackKhor:2000}, weak
\cite{HacKhoKri:2004} or off-diagonal admissibility \cite{ChaDew:2005,AmbDar:2013}.

\begin{define}[Block Tree] \label{def:blocktree}
  Let \(T_I, T_J\) be two cluster trees and let \(\op{adm} : T_I \times T_J \to \mathbbm{B} \). The \emph{block tree}
  \(T = T_{I \times J} \) is recursively defined starting with \( \troot{T} = (I,J) \):
  \begin{align*}
    & \sons{\clt,\cls} = \\
    & \begin{cases} \emptyset, \textnormal{ if } \op{adm}(\clt,\cls) = \textnormal{true} \vee
      \clt \in \mcL(T_I) \vee \cls \in \mcL(T_J),\\
      \set{ (\clt',\cls') \,:\, \clt' \in \sons{\clt}, \cls' \in \sons{\cls} } \textnormal{ else}.
    \end{cases}
  \end{align*}
  A node in \(T\) is also called a \emph{block}. Again, the set of leaves of \(T\) is denoted by \(\mcL(T) := \set{ b
    \in T \;:\; \sons{b} = \emptyset}\) and \(\level(\clt,\cls) = \level(\clt) = \level(\cls)\) denotes the distance
  from the root \(I \times J\).
\end{define}

If for a given block tree the rank of the low-rank approximation within admissible blocks can be bounded, one obtains
the set of \mcH-matrices:
\begin{define}[\mcH-Matrix]
  For a block tree $T$ over cluster trees $T_I, T_J$ and $k \in \N$, the set of \mcH-matrices $\mcH(T,k)$ is defined as
  \begin{align*}
    \mcH(T,k) := \{& M \in \R^{I \times J} \;:\; \forall (\clt,\cls) \in \mcL(T) : \\
                   & \rank(M_{\clt,\cls}) \le k \vee \clt \in \mcL(T_I) \vee \cls \in \mcL(T_J)\}
  \end{align*}
  Here, \(M_{\clt,\cls}\) refers to the sub-block \(M|_{\clt \times \cls}\).
\end{define}

In practice the constant rank \(k\) is typically replaced by a fixed low-rank approximation accuracy \(\varepsilon > 0\)
as the resulting \mcH-matrices are often more memory efficient. For this we assume for an admissible block
\(M_{\clt,\cls}\) and its low-rank approximation \(U_{\clt,\cls} V_{\clt,\cls}^H, U_{\clt,\cls} \in \R^{\clt \times k'},
V_{\clt,\cls} \in \R^{\cls \times k'}\):
\begin{equation} \label{eqn:epsilon}
||M_{\clt,\cls} - U_{\clt,\cls} V_{\clt,\cls}^H|| \le \varepsilon ||M_{\clt,\cls}|| \ .
\end{equation}
Here, \(k'\) is an \(\varepsilon\)-dependent rank.
This then leads to the definition of the set \(\mcH(T,\varepsilon)\) in an analogue way to \(\mcH(T,k)\) as the set of
\mcH-matrices with \emph{local} low-rank approximation error \(\varepsilon\). We will also use \(\mcH(T)\) if either
a fixed rank or a fixed accuracy is used.

\begin{remark}
  The set \(\mcH(T)\) also includes various other formats like \emph{block low-rank} (BLR) \cite{AmeAshWeis:2015} with a
  single hierarchy level or \emph{hierarchical off-diagonal low-rank} (HODLR) \cite{AmbDar:2013} with inadmissibility
  restricted to the diagonal, as only the clustering or the admissibility condition has to be chosen appropriately.
\end{remark}


\begin{figure}[htb]
  \centering
  \pgfdeclareimage[width=.24\textwidth]{memdim}{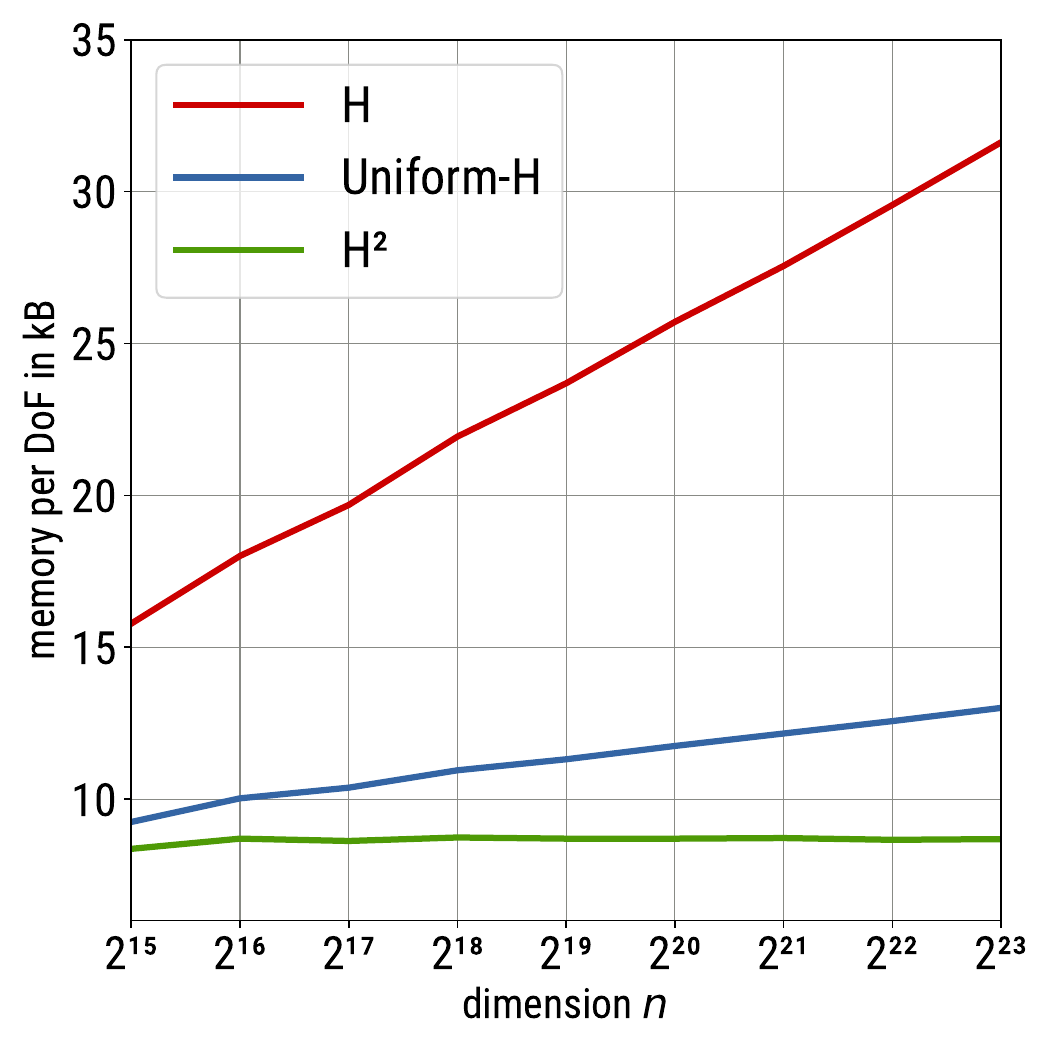}
  \pgfdeclareimage[width=.24\textwidth]{memeps}{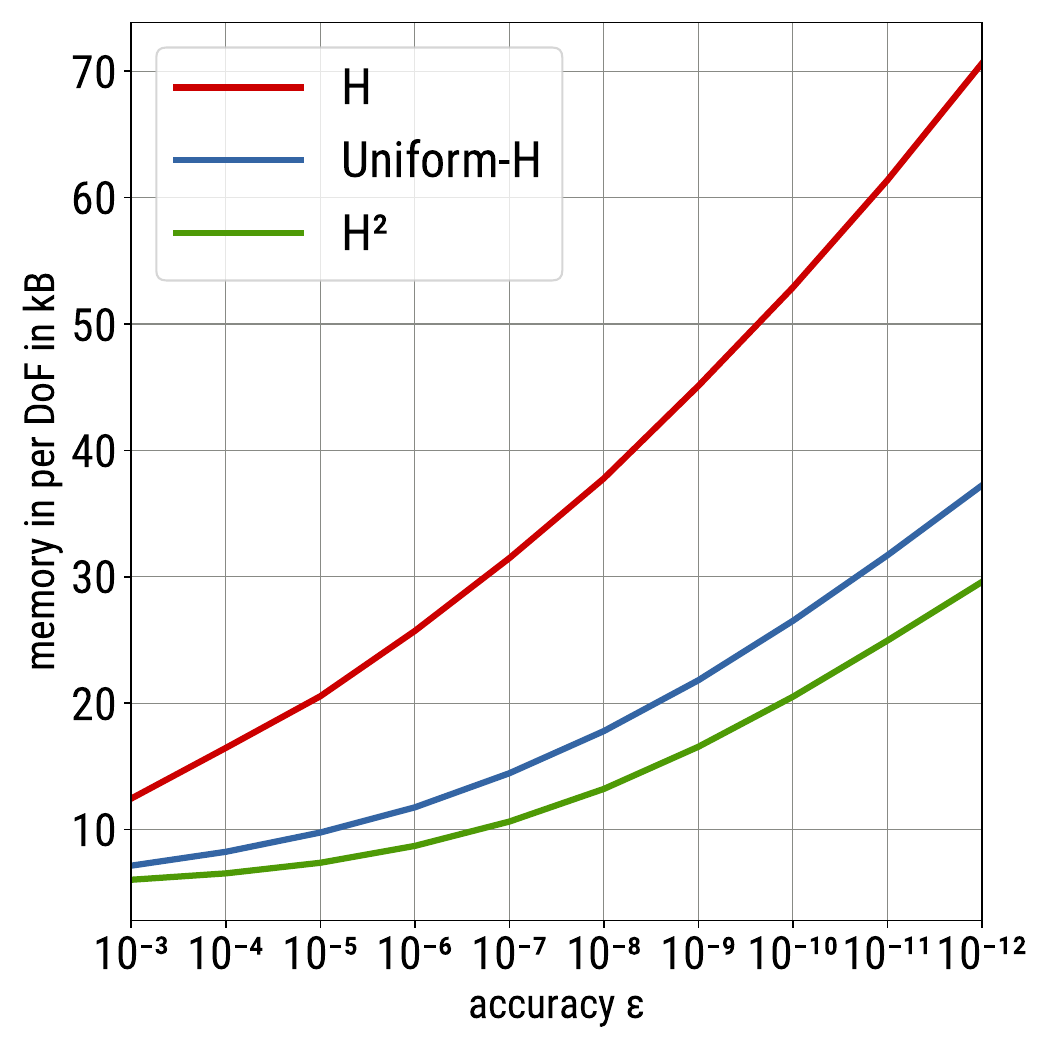}
  \pgfuseimage{memdim}\pgfuseimage{memeps}
  \caption{Matrix storage for different \mcH formats depending on matrix size (left) and accuracy (right).}
  \label{fig:mem}
\end{figure}

To simplify the access to certain sub sets of the matrix blocks of a given \mcH-matrix, some further sets are defined:
\begin{define}
  With \(\mcL(M) := \set{M_{\clt,\cls}\,:\,(\clt,\cls) \in \mcL(T)}\) we will denote the set of sub-blocks in \(M\)
  corresponding to leaves in \(T\). For \(\clt \in T_I\) the set \(\mcM^r_{\clt} := \set{M_{\clt,\cls} \in \mcL(M)}\)
  denotes the set of matrix block in the block row defined by \(\clt\). In an analogue way the set
  \(\mcM^c_{\cls} := \set{M_{\clt,\cls} \in \mcL(M)}\) for the block column is defined. The restriction of both sets to
  low-rank blocks shall be denoted by \(\mcM^{r,lr}_{\clt}\) and \(\mcM^{c,lr}_{\cls}\), respectively.
\end{define}


\begin{figure*}[htb]
  \centering
  \begin{tikzpicture}
    \begin{scope}
      \drawhmat
    \end{scope}
    \begin{scope}[xshift=5cm]
      \drawhhmat

      \begin{scope}[yshift=-4.25cm,yscale=-1]

        \foreach \ofs [evaluate=\ofs] in {0,0.5,...,3.75} {
          \draw (\ofs+0.125,0.0625) -- ++(0.125,0.4375cm);
          \draw (\ofs+0.375,0.0625) -- ++(-0.125,0.4375cm);
        }
        
        \begin{scope}[scale=0.25]
          \foreach \ofs in {0,1,...,15} {
            \draw [fill=coladm,opacity=0.5] (\ofs,0cm) rectangle ++(1cm,0.25cm);
          }
        \end{scope}
        
        \begin{scope}[yshift=0.5cm]
          \foreach \ofs [evaluate=\ofs] in {0,1,...,3} {
            \draw (\ofs+0.25,0.125) -- ++(0.25,0.375cm);
            \draw (\ofs+0.75,0.125) -- ++(-0.25,0.375cm);
          }
          \begin{scope}[scale=0.5]
            \foreach \ofs [evaluate=\ofs] in {0,1,...,7} {
              \draw [fill=coladm,opacity=0.5] (\ofs,0cm) rectangle ++(1cm,0.25cm);
            }
          \end{scope}
        \end{scope}

        \begin{scope}[yshift=1cm]
          \foreach \ofs [evaluate=\ofs] in {0,2} {
            \draw (\ofs+0.5,0.25) -- ++(0.5,0.375cm);
            \draw (\ofs+1.5,0.25) -- ++(-0.5,0.375cm);
          }
          \foreach \ofs [evaluate=\ofs] in {0,1,...,3} {
            \draw [fill=coladm,opacity=0.5] (\ofs,0cm) rectangle ++(1cm,0.25cm);
          }
        \end{scope}

        \begin{scope}[yshift=1.5cm]
          \draw (1,0.125) -- ++(1,0.375cm);
          \draw (3,0.125) -- ++(-1,0.375cm);
        \end{scope}
      \end{scope}
    \end{scope}
    \begin{scope}[xshift=10cm]
      \drawhhmat

      \begin{scope}[yshift=-4.25cm,yscale=-1]

        \foreach \ofs [evaluate=\ofs] in {0,0.5,...,3.75} {
          \draw (\ofs+0.125,0.0625) -- ++(0.125,0.4375cm);
          \draw (\ofs+0.375,0.0625) -- ++(-0.125,0.4375cm);
        }
        
        \begin{scope}[scale=0.25]
          \foreach \ofs in {0,1,...,15} {
            \draw [fill=coladm,opacity=0.5] (\ofs,0cm) rectangle ++(1cm,0.25cm);
          }
        \end{scope}
        
        \begin{scope}[yshift=0.5cm]
          \foreach \ofs [evaluate=\ofs] in {0,1,...,3} {
            \draw (\ofs+0.25,0.125) -- ++(0.25,0.375cm);
            \draw (\ofs+0.75,0.125) -- ++(-0.25,0.375cm);
          }
          \begin{scope}[scale=0.5]
            \foreach \ofs [evaluate=\ofs] in {0,1,...,7} {
              \draw [fill=colinadm] (\ofs+0.375,0cm) rectangle ++(0.25cm,0.25cm);
            }
          \end{scope}
        \end{scope}

        \begin{scope}[yshift=1cm]
          \foreach \ofs [evaluate=\ofs] in {0,2} {
            \draw (\ofs+0.5,0.25) -- ++(0.5,0.375cm);
            \draw (\ofs+1.5,0.25) -- ++(-0.5,0.375cm);
          }
          \foreach \ofs [evaluate=\ofs] in {0,1,...,3} {
            \draw [fill=colinadm] (\ofs+0.375,0cm) rectangle ++(0.25cm,0.25cm);
          }
        \end{scope}

        \begin{scope}[yshift=1.5cm]
          \draw (1,0.125) -- ++(1,0.375cm);
          \draw (3,0.125) -- ++(-1,0.375cm);
        \end{scope}
      \end{scope}
    \end{scope}
  \end{tikzpicture}
  \caption{Separate (left), shared (center) and nested (right) cluster bases for \mcH-matrices, \mcUH-matrices
    and \mcHH-matrices.}
  \label{fig:clbases}
\end{figure*}

\subsection{Uniform \mcH-Matrices}

In \mcH-matrices the low-rank blocks are represented in factored form \(U_{\clt,\cls} V_{\clt,\cls}^H\). Alternatively,
one can use \(\mcW_{\clt,\cls} S_{\clt,\cls} \mcX_{\clt,\cls} \) with orthogonal (row basis) \(\mcW_{\clt,\cls}\) and
(column basis) \(\mcX_{\clt,\cls}\) and a \emph{coupling} matrix \(S_{\clt,\cls} \in \R^{k \times k}\), e.g., by
computing a QR factorization of \(U_{\clt,\cls} = Q_U R_U\) and \(V_{\clt,\cls} = Q_V R_V\) and setting
\(\mcW_{\clt,\cls} := Q_U, \mcX_{\clt,\cls} := Q_V\) and \(S_{\clt,\cls} := R_U R_V^H\). In this case, each low-rank
block has individual \emph{row} and \emph{colum (cluster) bases}. However, for many applications one can find a
\emph{shared} cluster basis \(\mcW_{\clt}\) for all low-rank blocks \(M_{\clt,\cls'} \in \mcM^r_{\clt}\) in a single
block row. All such cluster bases form a tree \(T_{\mcW} := \set{ \mcW_{\clt}\;:\;\clt \in T_I}\) structurally identical
to the cluster tree \(T_I\). For simplicity we assume a cluster basis with rank 0 in case that no low-rank block exists
in the corresponding block row. In Figure~\ref{fig:clbases} (center) this can be seen on the two top levels of the
cluster bases tree. In an analogue way shared column bases \(\mcX_{\cls}, \cls \in T(J),\) and the corresponding tree
\(T_{\mcX}\) are defined.

A low-rank block \(M_{\clt,\cls}\) is then represented as
\begin{displaymath}
  M_{\clt,\cls} = \mcW_{\clt} \mcS_{\clt,\cls} \mcX_{\cls}^H .
\end{displaymath}
Such \mcH-matrices are called \emph{uniform} \mcH-matrices (\mcUH-matrices) and where first introduced together with
\mcH-matrices in \cite{Hackbusch:1999} and investigated in more detail in \cite{BruHuyMee:2025} where also algorithms
for constructing shared cluster bases are presented. While the storage for the coupling matrices, i.e., the actual
matrix data, is reduced to \landau{n} complexity, the overall memory complexity of \mcUH-matrices is
identical to (standard) \mcH-matrices due to \landau{n \log n} storage for the cluster bases trees \(T_{\mcW}\) and
\(T_{\mcX}\). However, in practice memory is often significantly smaller as is shown in Figure~\ref{fig:mem} where the
memory costs for the different matrix formats are presented. Especially noteworthy is the slower increase of the
storage for \mcUH-matrices compared to \mcH-matrices for an increasing problem size (Figure~\ref{fig:mem} (left)).


\subsection{\mcHH-Matrices}

Further reduction of the storage costs for the cluster bases can be achieved by exploiting the fact that for many
applications the bases have a \emph{nestedness} property, i.e., a cluster basis on an upper level can be represented by
a linear combination of bases on lower levels of the tree. For this, let \(\clt \in T_I\) with two child clusters:
\(\mcS(\clt) = \set{ \clt_0, \clt_1 }\). The cluster basis \(\mcW_{\clt}\) can then be represented as
\begin{displaymath}
  \mcW_{\clt} =
  \begin{pmatrix}
    \mcW_{\clt_0} E_{\clt,0} \\
    \mcW_{\clt_1} E_{\clt,1} 
  \end{pmatrix}
\end{displaymath}
using the cluster bases \(\mcW_{\clt_0}, \mcW_{\clt_1}\) from the child clusters and so-called \emph{transfer matrices}
\(E_{\clt,0}, E_{\clt,1} \in \R^{k \times k}\). Due to the recursive structure of the cluster basis tree, \(\mcW_{\clt_0}\) and
\(\mcW_{\clt_1}\) can again be represented by their corresponding child clusters in the same way until the leaves of the
tree are reached. Only there, cluster bases are stored explicitly as for \mcUH-matrices. For all other bases only
the transfer matrices are needed, reducing the per-basis storage from \(\landau{\clt}\) to \(\landau{k}\). This is shown
in Figure~\ref{fig:clbases} (right).

Representing \mcH-matrices by such \emph{nested} bases leads to \mcHH-matrices as introduced in \cite{HacKhoSau:2000}
and thoroughly discussed in \cite{Boerm:2010}. With this approach the storage complexity for the cluster basis and with
this for the full \mcHH-matrix is reduced to linear complexity, which is visible by the constant memory per degree of
freedom independent on the matrix size in Figure~\ref{fig:mem} (left).


\section{Matrix-Vector Multiplication} \label{sec:mvm}

In the following the matrix-vector multiplication for the different formats of hierarchical matrices shall be
discussed for the update operation
\begin{equation} \label{eqn:mvm}
  y := \alpha M x + y
\end{equation}
with an hierarchical matrix \(M\) and vectors \(x\) and \(y\). The special focus is here on the parallelization of the
various algorithms.

\begin{remark}
  All numerical experiments are performed on a single AMD Epyc 9554 CPU with 64 cores and 12 32GB DDR5-4800 memory
  modules, thereby maximizing the memory throughput of this processor with 12 memory channels. For parallelization Intel
  TBB v2022.1 was used while Intel oneMKL v2025.1 (sequential version, AVX512 code path) provided the BLAS and LAPACK
  functions for the uncompressed case. All code was compiled using GCC v14.2. The algorithms described in this work are
  implemented in the open source software HLR (version \emph{c8810d5b})\footnote{\url{http://libhlr.org}, programs:
    \emph{mvm, mvm-uni} and \emph{mvm-h2}}.
\end{remark}

\subsection{\mcH-Matrices}

The product \eqref{eqn:mvm} is computed by iterating over the leaf blocks of \(M\) and performing local matrix-vector
multiplications, either with a dense matrix for inadmissible blocks or in low-rank format, i.e.,
\(t := V_{\clt,\cls}^H x|_{\cls}\) followed by \(y|_{\clt} := y|_{\clt} + \alpha U_{\clt,\cls} t\). The full procedure
is shown in Algorithm~\ref{alg:hmvm}.

\begin{algorithm}{\mcH-Matrix-Vector Multiplication}{alg:hmvm}
  \Procedure{hmvm}{$\alpha,M,x,y$}
  \For{\((\clt,\cls) \in \mcL(M)\)}
    \If{\((\clt,\cls\)) is admissible}
      \State \(t := \alpha U_{\clt,\cls} V_{\clt,\cls}^H x|_{\cls}\);
    \Else
      \State \(t := \alpha D_{\clt,\cls} x|_{\cls}\);
    \EndIf 
    \State \(y|_{\clt} := y|_{\clt} + t\);
  \EndFor
  \EndProcedure
\end{algorithm}

Versions of the \mcH-MVM for parallel systems need to consider load balancing due to different,
not a priori known ranks in different low-rank blocks of the \mcH-matrix if a fixed accuracy \(\varepsilon\) is used. Also
the block structure is typically not equal throughout the matrix. This poses a serious scalability issue for the
distributed memory case (see \cite{BebKri:2005,LiPouYin:2021}) or systems with a NUMA architecture.

On shared memory systems a task-based approach can avoid these problems if the scheduling algorithm is able to assign
ready tasks efficiently to idle processors. However, this may lead to other problems as the memory layout of the blocks handled by a
single processor may not be optimal for efficient execution. This is of special importance because of the low
computational intensity of matrix-vector multiplication, which normally leads to a memory bandwidth limited
performance. Different optimization strategies are discussed in \cite{HosIdaHan:2022}, where especially the memory
layout of the \mcH-matrix data is adjusted such that memory loads are faster.

Another issue with shared memory programming is handling potential collisions when writing to the same memory positions,
e.g., with matrix blocks \(M_{\clt,\cls}\) and \(M_{\clt,\cls'}\) handled by different processors (or processor cores)
writing simultaneously to \(y|_{\clt}\). In the literature, different solutions to this problem can be found. In \cite{IdaIwaMifTak:2014}
\emph{atomic} updates to vector coefficients are used. The implementation in \cite{HLIBpro} splits the vector \(y\) into
\emph{chunks} \(y|_{\clt}, \clt \in \mcL(T_I)\). Updates to \(y\) are then performed for each such chunk
seperately. Data integrity for parallel updates is ensured by a mutex for each \(y|_{\clt}\). This is implemented in
Algorithm~\ref{alg:chunks} which replaces the direct update in the last line of Algorithm~\ref{alg:hmvm}.

\begin{algorithm}{Chunk based Updates}{alg:chunks}
  \Procedure{update}{$y,\clt,t$}
  \If{\(\clt \in \mcL(T_I)\)}
    \State lock \function{mutex}[$y|_{\clt}$]
    \State \(y|_{\clt} := y|_{\clt} + t|_{\clt}\);
    \State unlock \function{mutex}[$y|_{\clt}$]
  \Else
    \ForAll{\(\clt' \in \mcS(\clt)\)}
      \State \function{update}[$y,\clt',t$]
    \EndFor
  \EndIf
  \EndProcedure
\end{algorithm}

Atomic updates and mutexes allow the matrix-vector multiplication to be performed in parallel on the full set of dense
and low-rank matrix blocks but depend on the task scheduling scheme for reducing collisions and may suffer from
potential performance problems with an increasing number of processor cores. Here, a reduction approach of (thread)
local results as is used in \cite{BebKri:2005,LiPouYin:2021} may eliminate these problems but again increases the
overall memory usage due to multiple copies of the vector data and higher computational costs due to the additional
reduction phase, especially for processors with many cores.

\begin{figure}[htb]
  \centering
  \begin{tikzpicture}[xscale=0.6,yscale=-0.6]
    \begin{scope}[scale=0.5]
      \draw (0,0) grid ++(8,8);
    \end{scope}

    \begin{scope}[scale=0.5]
      \foreach \x/\y in {1,2,3,4,5,6,7} {
        \draw [fill=SkyBlue3,opacity=0.5] (\x,0) rectangle ++(0.25,1);
        \draw [fill=Chameleon3,opacity=0.5] (\x,0) rectangle ++(1,0.25);
      }
      \foreach \x/\y in {0,2,3,4,5,6,7} {
        \draw [fill=SkyBlue3,opacity=0.5] (\x,1) rectangle ++(0.25,1);
        \draw [fill=Chameleon3,opacity=0.5] (\x,1) rectangle ++(1,0.25);
      }
      \foreach \x/\y in {0,1,3,4,5,6,7} {
        \draw [fill=SkyBlue3,opacity=0.5] (\x,2) rectangle ++(0.25,1);
        \draw [fill=Chameleon3,opacity=0.5] (\x,2) rectangle ++(1,0.25);
      }
      \foreach \x/\y in {0,1,2,4,5,6,7} {
        \draw [fill=SkyBlue3,opacity=0.5] (\x,3) rectangle ++(0.25,1);
        \draw [fill=Chameleon3,opacity=0.5] (\x,3) rectangle ++(1,0.25);
      }
      \foreach \x/\y in {0,1,2,3,5,6,7} {
        \draw [fill=SkyBlue3,opacity=0.5] (\x,4) rectangle ++(0.25,1);
        \draw [fill=Chameleon3,opacity=0.5] (\x,4) rectangle ++(1,0.25);
      }
      \foreach \x/\y in {0,1,2,3,4,6,7} {
        \draw [fill=SkyBlue3,opacity=0.5] (\x,5) rectangle ++(0.25,1);
        \draw [fill=Chameleon3,opacity=0.5] (\x,5) rectangle ++(1,0.25);
      }
      \foreach \x/\y in {0,1,2,3,4,5,7} {
        \draw [fill=SkyBlue3,opacity=0.5] (\x,6) rectangle ++(0.25,1);
        \draw [fill=Chameleon3,opacity=0.5] (\x,6) rectangle ++(1,0.25);
      }
      \foreach \x/\y in {0,1,2,3,4,5,6} {
        \draw [fill=SkyBlue3,opacity=0.5] (\x,7) rectangle ++(0.25,1);
        \draw [fill=Chameleon3,opacity=0.5] (\x,7) rectangle ++(1,0.25);
      }
    \end{scope}
    

    \draw [line width=2pt,->] (4.5,2) -- (5.5,2);
    
    \begin{scope}[xshift=5cm]
      \begin{scope}[scale=0.5,xshift=2cm]
        \foreach \y in {0,1,2,3,4,5,6,7} {
          \foreach \x in {0,0.25,...,1.5} {
            \draw [fill=SkyBlue3,opacity=0.5] (\x,\y) rectangle ++(0.25,1);
          }
        }
      \end{scope}
    \end{scope}
    \begin{scope}[xshift=8cm]
      \begin{scope}[scale=0.5]
        \foreach \x in {0,1,2,3,4,5,6,7} {
          \foreach \y in {0,0.25,...,1.5} {
            \draw [fill=Chameleon3,opacity=0.5] (\x,\y) rectangle ++(1,0.25);
          }
        }
      \end{scope}
    \end{scope}
  \end{tikzpicture}
  \caption{Stacking of low-rank factors for BLR clustering.}
  \label{fig:stackedblr}
\end{figure}

An alternative approach is a collision free design in which the memory blocks are scheduled to the processors in a way
to prevent simultaneous writing to the same memory positions. For the (non-hierarchical) BLR format with its \(p \times
q\) set of dense and low-rank sub-blocks this is naturally achieved by performing the matrix-vector multiplication for
all blocks in a single block row in a single task. In \cite{LtaCraGra:2021} this is further optimized by stacking the
local low-rank factors of these sub-blocks together as is shown in Figure~\ref{fig:stackedblr}. With this, multiple
sub-block matrix-vector multiplications are combined into a single multiplication with an additional exchange of
intermediate data. This approach can also be applied to \mcH-matrices where the low-rank factors of sub-blocks on the
\emph{same level} are stacked together (see Figure~\ref{fig:stackeduv}).

\begin{figure}[htb]
  \centering
  \begin{tikzpicture}[xscale=0.6,yscale=-0.6]
    \draw (0,0) grid ++(4,4);

    \begin{scope}[scale=0.5]
      \draw (0,0) grid ++(4,4);
      \draw (2,2) grid ++(4,4);
      \draw (4,4) grid ++(4,4);
    \end{scope}

    \foreach \x/\y in {2/0,3/0,3/1,0/2,0/3,1/3} {
      \draw [fill=SkyBlue1,opacity=0.5] (\x,\y) rectangle ++(0.25,1);
      \draw [fill=Chameleon1,opacity=0.5] (\x,\y) rectangle ++(1,0.25);
    }

    \begin{scope}[scale=0.5]
      \foreach \x/\y in {2/0,3/0,3/1,0/2,0/3,1/3} {
        \draw [fill=SkyBlue3,opacity=0.5] (\x,\y) rectangle ++(0.25,1);
        \draw [fill=Chameleon3,opacity=0.5] (\x,\y) rectangle ++(1,0.25);
      }
      \begin{scope}[xshift=2cm,yshift=2cm]
        \foreach \x/\y in {2/0,3/0,3/1,0/2,0/3,1/3} {
          \draw [fill=SkyBlue3,opacity=0.5] (\x,\y) rectangle ++(0.25,1);
          \draw [fill=Chameleon3,opacity=0.5] (\x,\y) rectangle ++(1,0.25);
        }
      \end{scope}
      \begin{scope}[xshift=4cm,yshift=4cm]
        \foreach \x/\y in {2/0,3/0,3/1,0/2,0/3,1/3} {
          \draw [fill=SkyBlue3,opacity=0.5] (\x,\y) rectangle ++(0.25,1);
          \draw [fill=Chameleon3,opacity=0.5] (\x,\y) rectangle ++(1,0.25);
        }
      \end{scope}
    \end{scope}
    

    \draw [line width=2pt,->] (4.5,2) -- (5.5,2);
    
    \begin{scope}[xshift=6cm]
      \foreach \y/\x in {0/0,0/0.25,1/0,2/0,3/0,3/0.25} {
        \draw [fill=SkyBlue1,opacity=0.5] (\x,\y) rectangle ++(0.25,1);
      }
      \begin{scope}[scale=0.5,xshift=2cm]
        \foreach \y/\x in {0/0,0/0.25,1/0,2/0,2/0.25,2/0.5,3/0,3/0.25,3/0.5,4/0,4/0.25,4/0.5,5/0,5/0.25,5/0.5,6/0,7/0,7/0.25} {
          \draw [fill=SkyBlue3,opacity=0.5] (\x,\y) rectangle ++(0.25,1);
        }
      \end{scope}
    \end{scope}
    \begin{scope}[xshift=8cm]
      \foreach \x/\y in {0/0,0/0.25,1/0,2/0,3/0,3/0.25} {
        \draw [fill=Chameleon1,opacity=0.5] (\x,\y) rectangle ++(1,0.25);
      }
      \begin{scope}[scale=0.5,yshift=2cm]
        \foreach \x/\y in {0/0,0/0.25,1/0,2/0,2/0.25,2/0.5,3/0,3/0.25,3/0.5,4/0,4/0.25,4/0.5,5/0,5/0.25,5/0.5,6/0,7/0,7/0.25} {
          \draw [fill=Chameleon3,opacity=0.5] (\x,\y) rectangle ++(1,0.25);
        }
      \end{scope}
    \end{scope}
  \end{tikzpicture}
  \caption{Stacking of low-rank factors per level of the \mcH-matrix.}
  \label{fig:stackeduv}
\end{figure}

A disadvantage of such a stacking method is that the data of low-rank blocks is no longer separate from each other.
This makes other \mcH-matrix computations, e.g., updates during an \mcH-LU factorization, more difficult especially
with non-constant ranks. Furthermore, the intermediate results of the multiplication with the \(V\) factors need
to be reassigned to fit the multiplication with the \(U\) factors, which induces additional overhead.

Therefore, the original data layout shall be kept while still reordering the sub-block multiplications such that data
access collisions are avoided. The basic idea is again to handle matrix-vector multiplication in a single block row
sequentially while following the \mcH-matrix hierarchy from the root to the leaf blocks.

Let \(\mcM^r := \set{ \mcM^r_{\clt}\;:\: \clt \in T_I}\) be the set of all block row sets. Since \(\mcM^r\) is defined
based on \(T_I\), it can be considered to be structurally identical to the cluster tree.
Figure~\ref{fig:clsort} shows this level- and cluster-wise sorting for a simple \mcH-matrix structure.

\begin{figure}[htb]
  \centering
  \begin{tikzpicture}[xscale=0.6,yscale=-0.6]
    \draw (0,0) grid ++(4,4);

    \begin{scope}[scale=0.5]
      \draw (0,0) grid ++(4,4);
      \draw (2,2) grid ++(4,4);
      \draw (4,4) grid ++(4,4);
    \end{scope}

    \foreach \x/\y in {2/0,3/0,3/1,0/2,0/3,1/3} {
      \draw [fill=SkyBlue1,opacity=0.5] (\x,\y) rectangle ++(0.25,1);
      \draw [fill=Chameleon1,opacity=0.5] (\x,\y) rectangle ++(1,0.25);
    }

    \begin{scope}[scale=0.5]
      \foreach \x/\y in {2/0,3/0,3/1,0/2,0/3,1/3} {
        \draw [fill=SkyBlue3,opacity=0.5] (\x,\y) rectangle ++(0.25,1);
        \draw [fill=Chameleon3,opacity=0.5] (\x,\y) rectangle ++(1,0.25);
      }
      \begin{scope}[xshift=2cm,yshift=2cm]
        \foreach \x/\y in {2/0,3/0,3/1,0/2,0/3,1/3} {
          \draw [fill=SkyBlue3,opacity=0.5] (\x,\y) rectangle ++(0.25,1);
          \draw [fill=Chameleon3,opacity=0.5] (\x,\y) rectangle ++(1,0.25);
        }
      \end{scope}
      \begin{scope}[xshift=4cm,yshift=4cm]
        \foreach \x/\y in {2/0,3/0,3/1,0/2,0/3,1/3} {
          \draw [fill=SkyBlue3,opacity=0.5] (\x,\y) rectangle ++(0.25,1);
          \draw [fill=Chameleon3,opacity=0.5] (\x,\y) rectangle ++(1,0.25);
        }
      \end{scope}
    \end{scope}
    

    \draw [line width=2pt,->] (4.5,2) -- (5.5,2);

    \begin{scope}[xshift=6cm]
      \foreach \y in {1,...,3} {
        \draw [Aluminium5,densely dotted] (-0.25,\y) -- ++(2.5,0);
      }
      \foreach \x/\y in {0/0,1/0,0/1,0/2,0/3,1/3} {
        \draw [fill=SkyBlue1,opacity=0.5] (\x,\y) rectangle ++(0.25,1);
        \draw [fill=Chameleon1,opacity=0.5] (\x,\y) rectangle ++(1,0.25);
        \draw (\x,\y) rectangle ++(1,1);
      }
    \end{scope}

    \begin{scope}[xshift=9cm,scale=0.5]
      \foreach \y in {1,...,7} {
        \draw [Aluminium5,densely dotted] (-0.5,\y) -- ++(5.5,0);
      }
      \foreach \x/\y in {0/0,1/0, 0/1, 0/2,1/2,2/2, 0/3,1/3,2/3,3/3, 0/4,1/4,2/4,3/4, 0/5,1/5,2/5,3/5, 0/6, 0/7,1/7} {
        \draw [fill=SkyBlue3,opacity=0.5] (\x,\y) rectangle ++(0.25,1);
        \draw [fill=Chameleon3,opacity=0.5] (\x,\y) rectangle ++(1,0.25);
        \draw (\x,\y) rectangle ++(1,1);
      }
    \end{scope}
  \end{tikzpicture}
  \caption{Sorting of \mcH-matrix blocks per block row and hierarchy level.}
  \label{fig:clsort}
\end{figure}

Now let \(\clt_0,\ldots,\clt_{\ell}\) be clusters of \(T_I\) with identical level, i.e., \(\level(\clt_i) =
\level(\clt_j), 0 \le i,j \le \ell\). Then, for any \(0 \le i,j \le \ell\) the matrix-vector products in the
corresponding sets \(\mcM^r_{\clt_i}\) and \(\mcM^r_{\clt_j}\) can be computed in parallel since \(\clt_i \cap \clt_j =
\emptyset\).

For any \(\clt,\cls \in T_I\) with \(\level(\clt) \ne \level(\cls)\) the sets \(\mcM^r_{\clt}\) and \(\mcM^r_{\cls}\) can only be
handled in parallel if \(\clt \cap \cls = \emptyset\). However, due to the definition of \(T_I\) if
\(\clt \cap \cls \ne \emptyset\) then either \(\clt \subseteq \cls\) or \(\cls \subseteq \clt\) holds. Therefore, if
\(T_I\) is traversed from root to leaves with execution of matrix blocks in a given \(\mcM^r_{\clt}\) before proceeding to
the child clusters in \(\mcS(\clt)\), any race condition when accessing \(y\) is prevented. This procedure is implemented in
Algorithm~\ref{alg:phmvm}.

\begin{algorithm}{Parallel \mcH-Matrix-Vector Multiplication}{alg:phmvm}
  \Procedure{phmvm}{$\alpha,\clt,\mcM^r,x,y$}
    \ForAll{\(M_{\clt,\cls} \in \mcM^r_{\clt}\)}
      \If{\(\clt,\cls\) is admissible}
        \State \(y|_{\clt} := y|_{\clt} + \alpha U_{\clt,\cls} V_{\clt,\cls}^H x|_{\cls}\);
      \Else
        \State \(y|_{\clt} := y|_{\clt} + \alpha D_{\clt,\cls} x|_{\cls}\);
      \EndIf 
    \EndFor
    \ParFor{\(\clt' \in \mcS(\clt)\)}
      \State \function{phmvm}[\(\alpha,\clt',\mcM^r,x,y\)]
    \EndFor
  \EndProcedure
\end{algorithm}

\begin{remark}
  Computing the adjoint product \(M^H x\) works in an analogue way by iterating over the corresponding block columns.
\end{remark}

\begin{remark}
  The stacked method for \mcH-matrices also needs to follow a root-to-leaf approach to prevent collisions when
  applying the vector update for non-disjoint clusters.
\end{remark}



In principle, the computation of all products for matrix blocks in \(\mcM^r_{\clt}\) in Algorithm~\ref{alg:phmvm} can be
further parallelized using a reduction scheme. One could also combine memory layout optimizations from
\cite{HosIdaHan:2022} easily with this approach. However, as no further improvement was observed in practice, such
techniques where not used in this work.

In Algorithm~\ref{alg:phmvm} no parallelism of the per-matrix-block products, e.g., of \(V_{\clt,\cls}^H \cdot
x|_{\cls}\) or the product with \(U_{\clt,\cls}\), was considered, which is a major drawback of this procedure if a
block structure like HODLR \cite{AmbDar:2013} is used. There, the most time consuming computations are performed on the
upper levels of the block cluster tree, where only a few processors may be used in parallel. However, as the numerical
results below demonstrate, typical \mcH-matrix block structures do not show these problems, at least not for the number
of processors cores considered in this work. Furthermore, such additional parallelization may be easily added to
Algorithm~\ref{alg:phmvm} if needed.

\begin{figure*}[htb]
  \centering
  \pgfdeclareimage[width=.31\textwidth]{mvmdH}{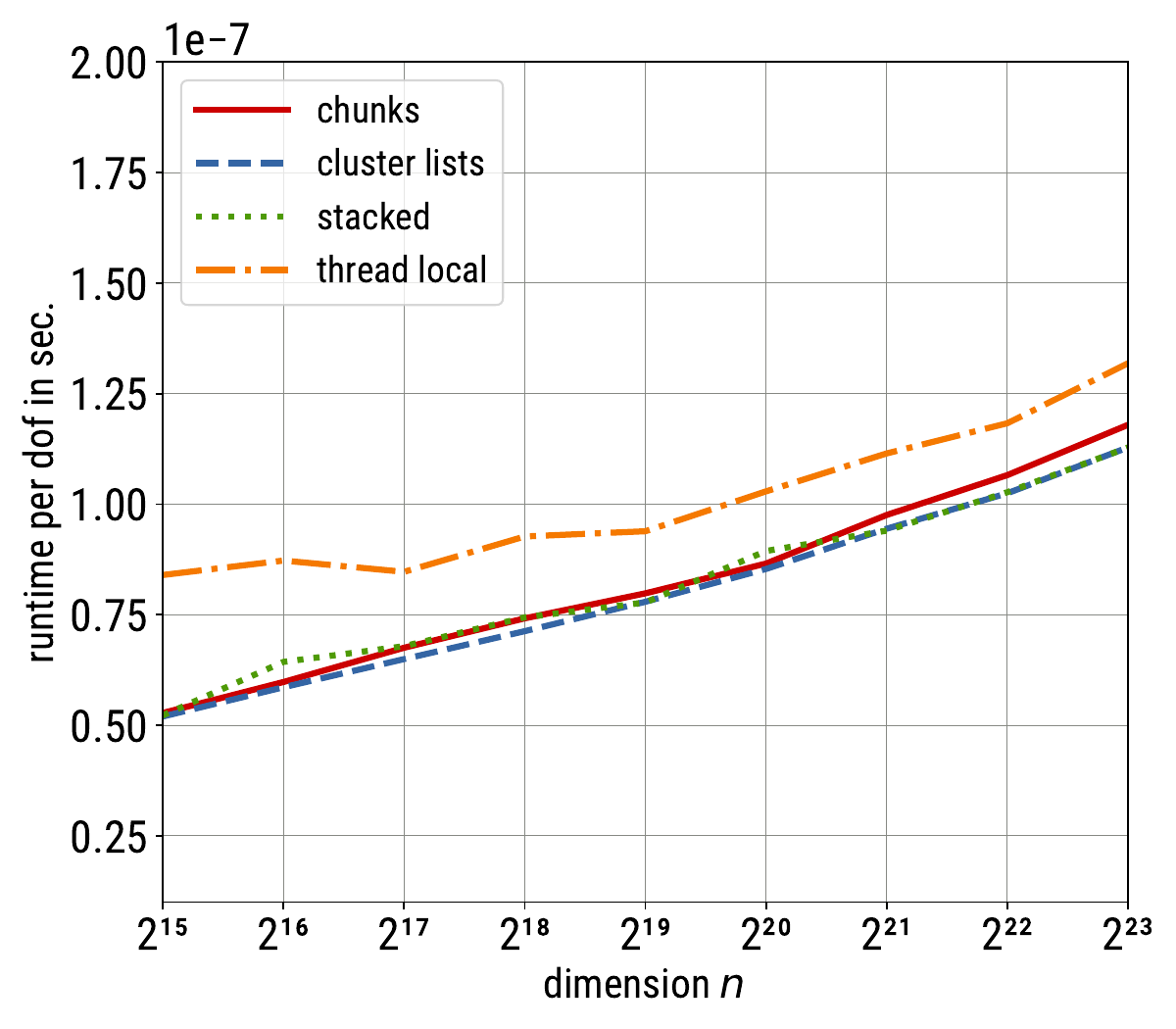}
  \pgfdeclareimage[width=.31\textwidth]{mvmdU}{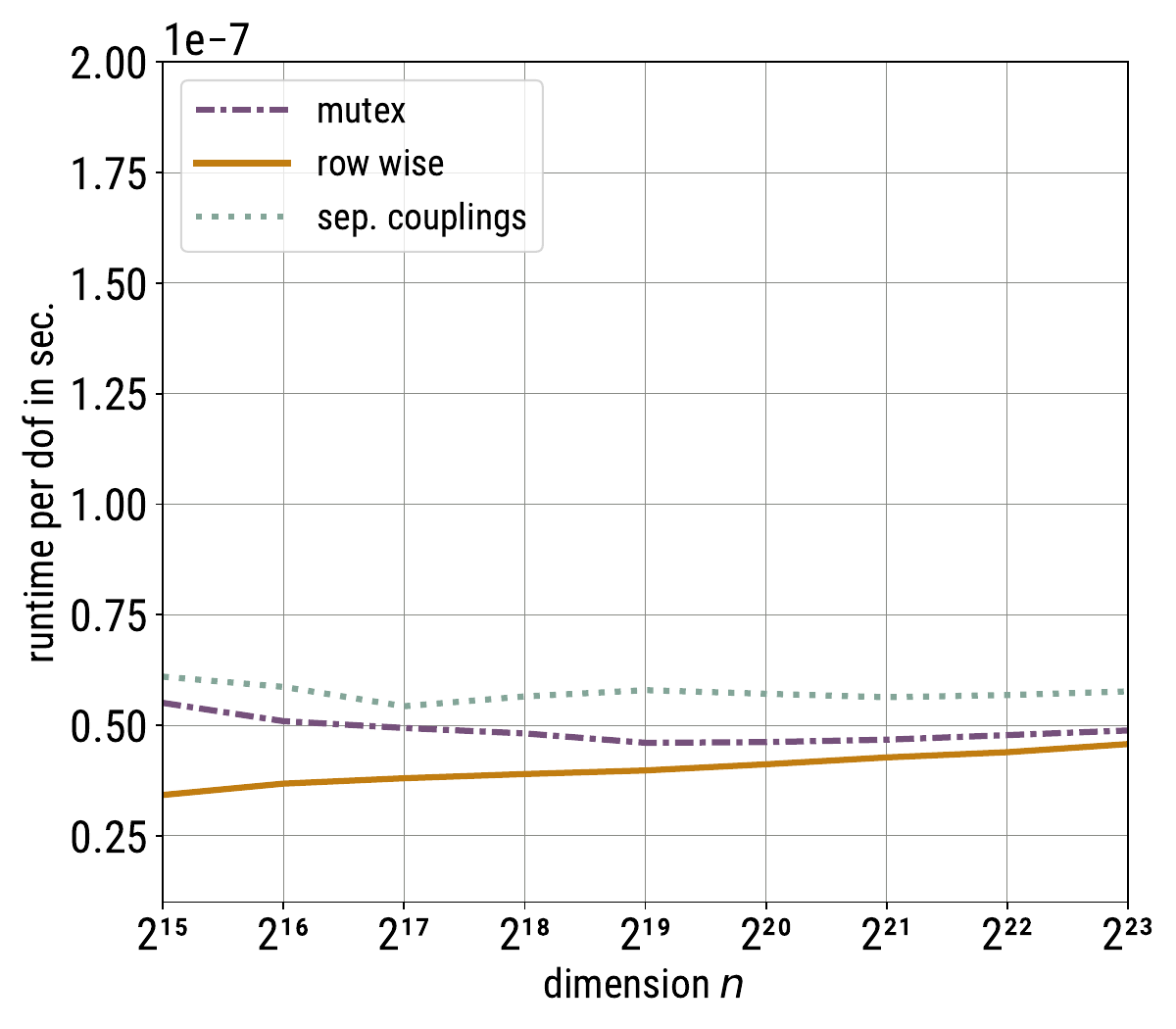}
  \pgfdeclareimage[width=.31\textwidth]{mvmd2}{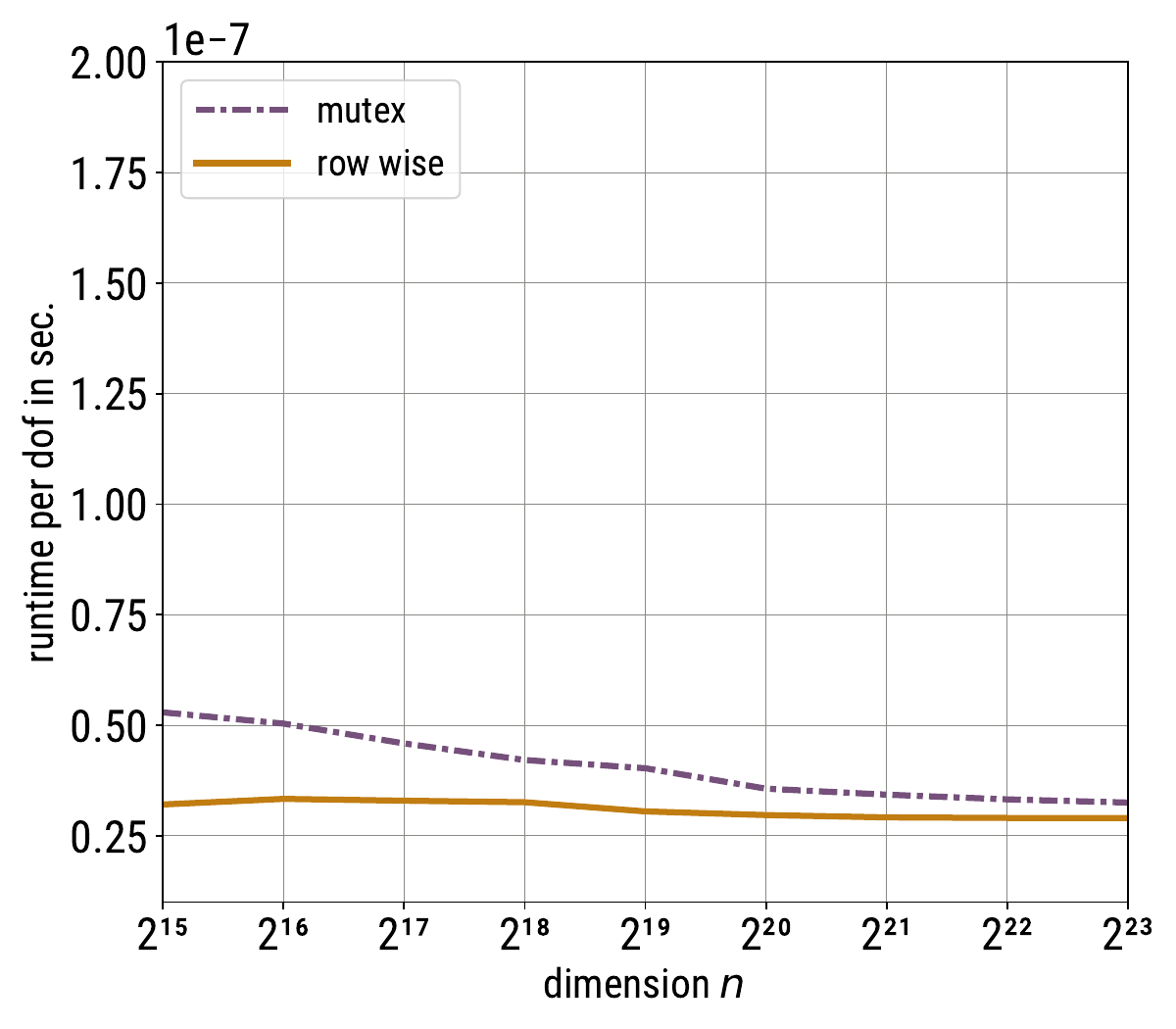}
  \pgfdeclareimage[width=.31\textwidth]{mvmeH}{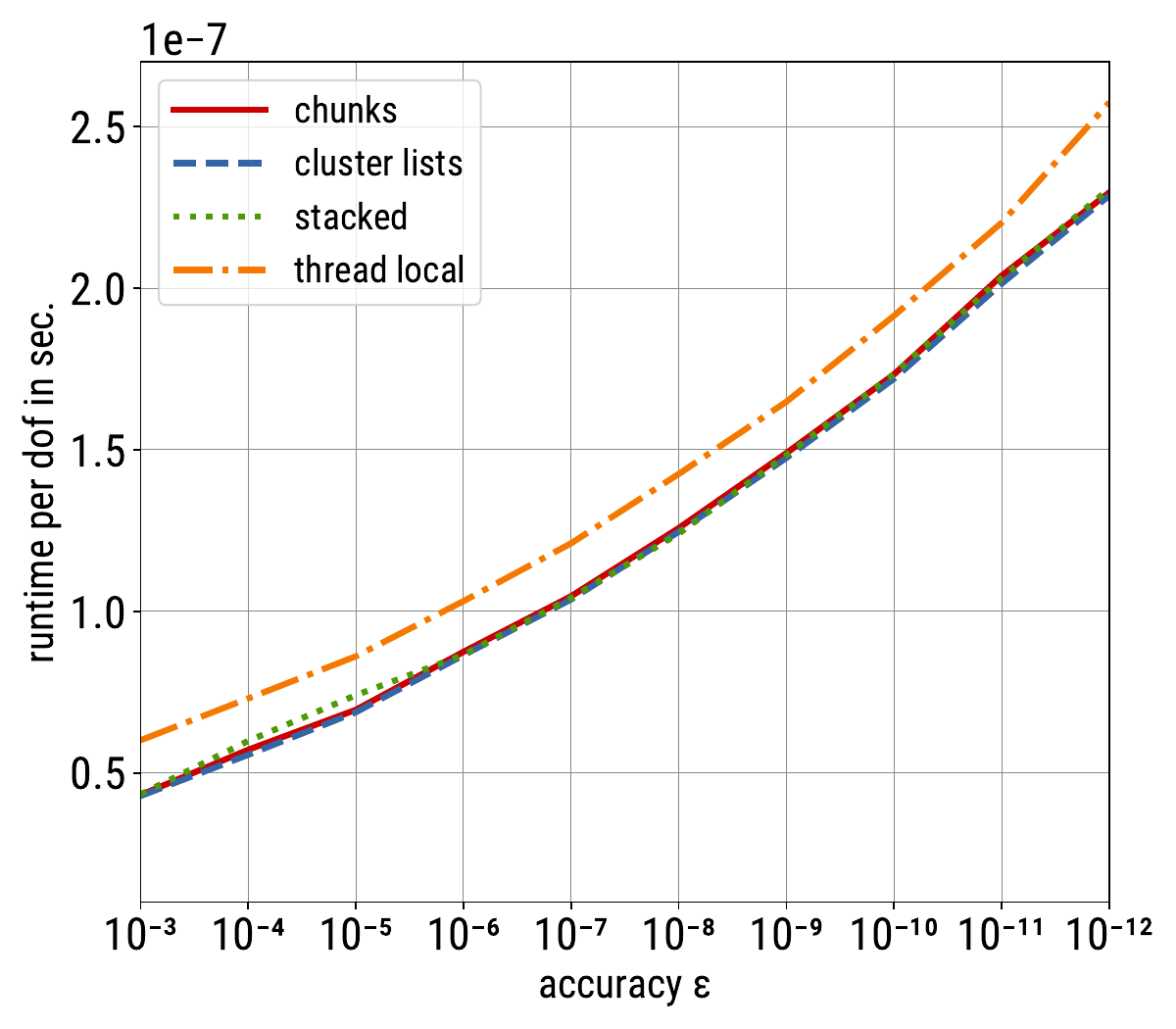}
  \pgfdeclareimage[width=.31\textwidth]{mvmeU}{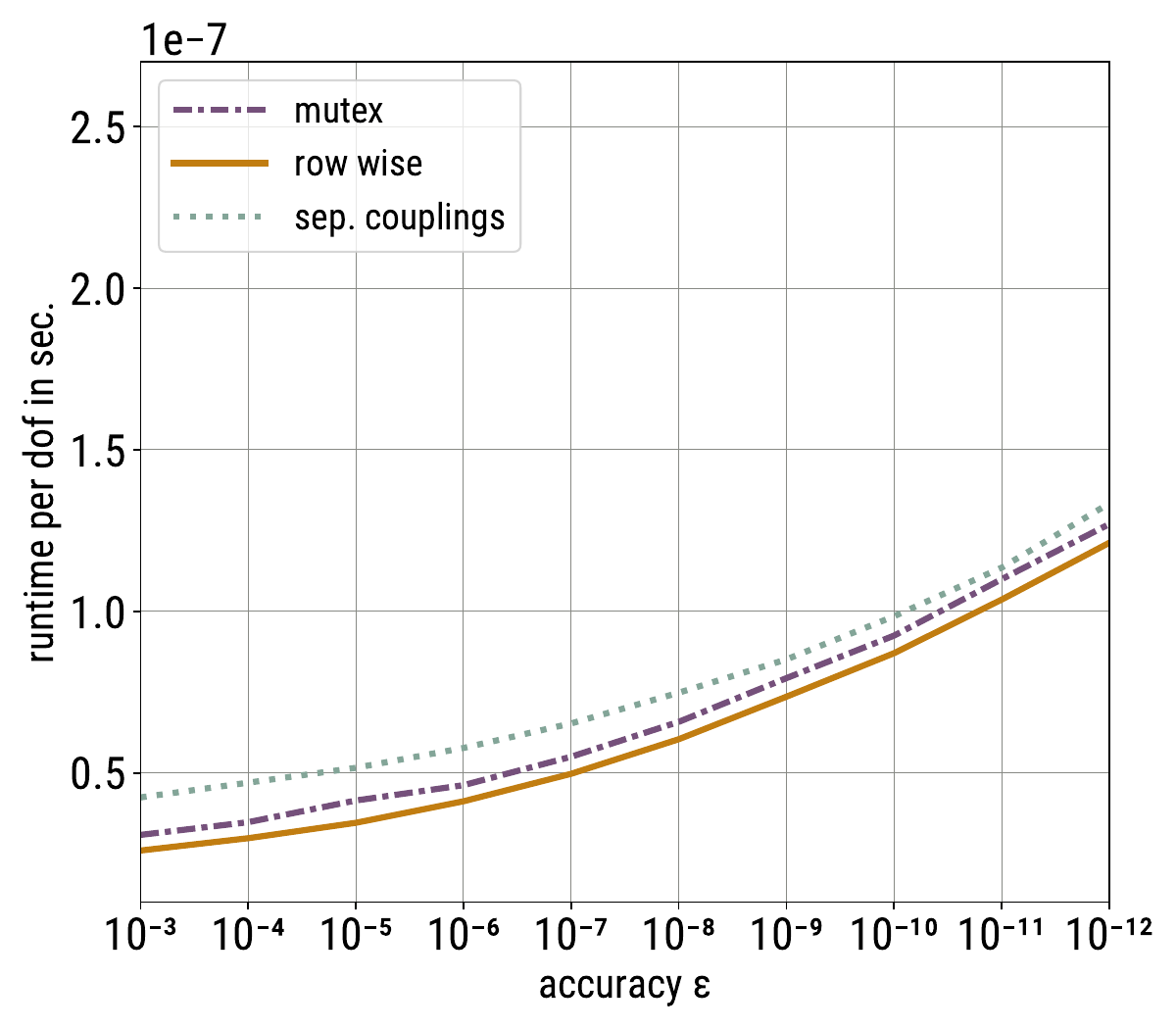}
  \pgfdeclareimage[width=.31\textwidth]{mvme2}{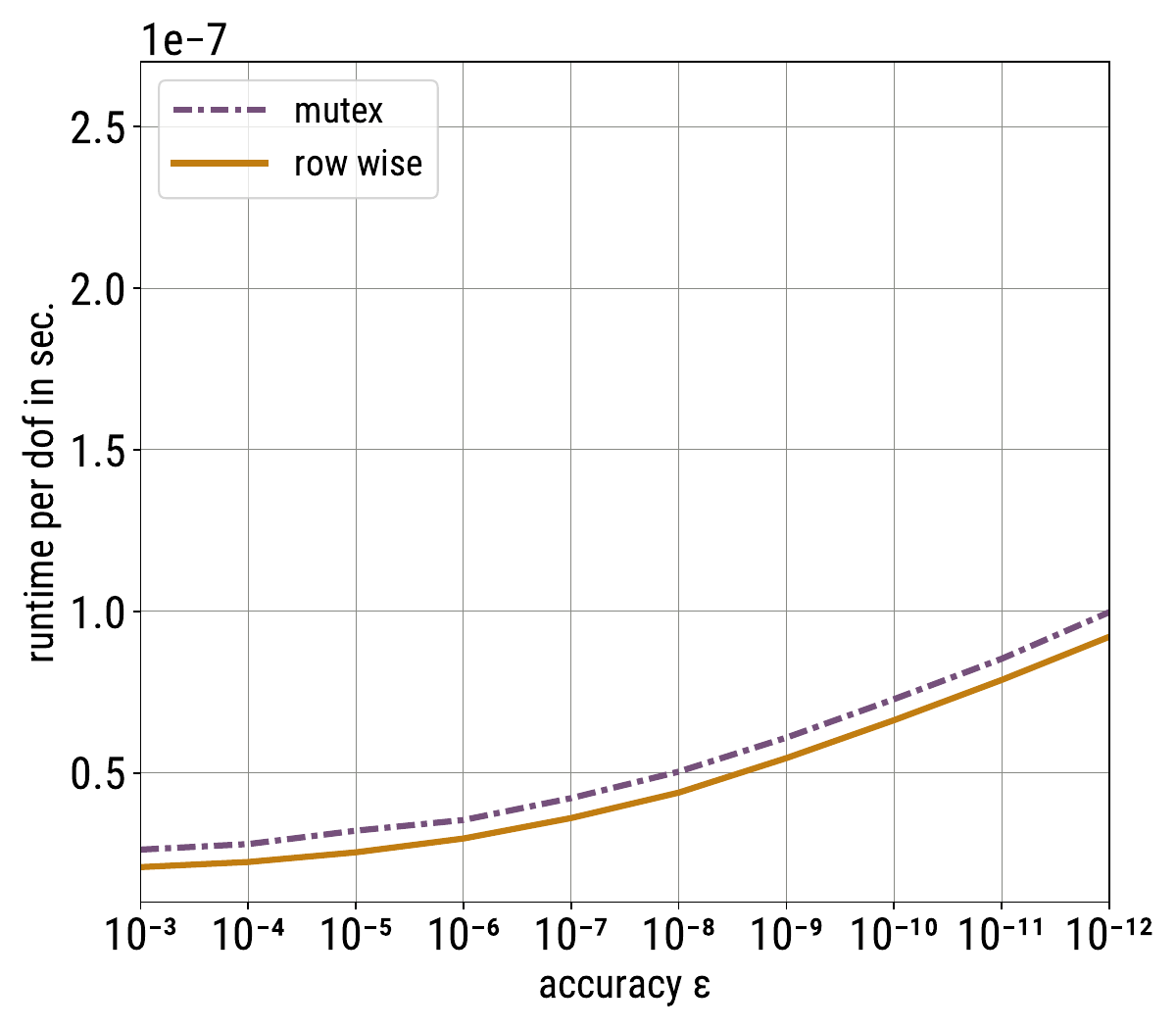}
  \begin{tabular}{ccc}
    \mcH & \mcUH & \mcHH \\
  \pgfuseimage{mvmdH} & \pgfuseimage{mvmdU} & \pgfuseimage{mvmd2}\\
  \pgfuseimage{mvmeH} & \pgfuseimage{mvmeU} & \pgfuseimage{mvme2}
  \end{tabular}
  \caption{Runtime of different matrix-vector multiplication algorithms for \mcH (left), \mcUH (center) and
    \mcHH-matrices (right).}
  \label{fig:mvmtime}
\end{figure*}

The runtime performance for the different \mcH-matrix-vector multiplication algorithms is shown in
Figure~\ref{fig:mvmtime} (left) for different problem sizes and accuracies. There, ``chunks'' denotes the mutex based
version from \cite{HLIBpro}, ``cluster lists'' refers to Algorithm~\ref{alg:phmvm}, the adaptation of the stacked local
cluster bases (low-rank factors) for \mcH-matrices shown as ``stacked'' and ``thread local'' uses a matrix-vector
multiplication with thread local results which are joined together afterwards.

As can be seen, the runtime of the multiplication using stacked cluster bases is identical to
Algorithm~\ref{alg:phmvm}. Almost identical results are achieved by using mutexes for leaf clusters, indicating that the
task scheduling is reasonably efficient on the given system. Only the usage of thread-local vectors inreases the
runtime, which is mostly due to the additional summation of the thread-local results. Leaving this part out reduces the
runtime almost to the same level as for the other methods.

\begin{figure}[htb]
  \centering
  \pgfdeclareimage[width=.43\textwidth]{rooflineH}{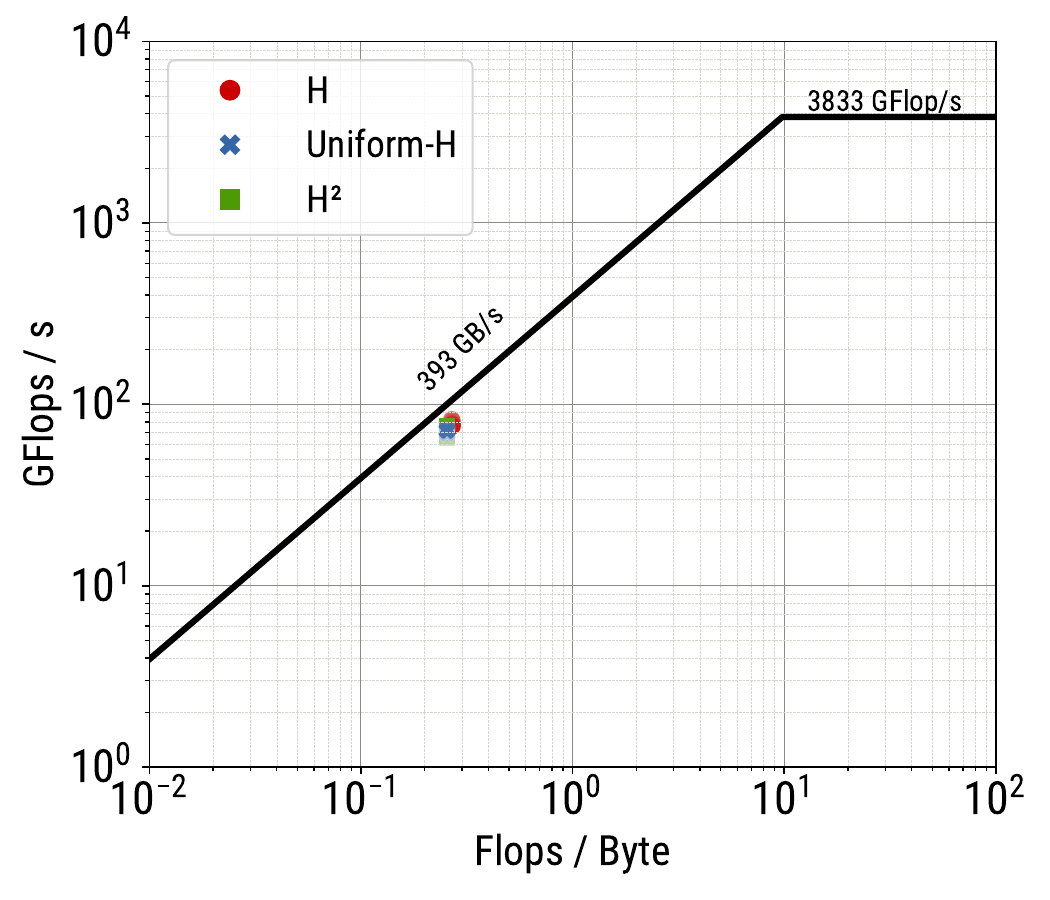}
  \pgfuseimage{rooflineH}
  \caption{Roofline plot for \mcH-MVM, \mcUH-MVM and \mcHH-MVM.}
  \label{fig:rooflineH}
\end{figure}

Furthermore, Algorithm~\ref{alg:phmvm}, bounded by the memory bandwidth, achieves 79\% of the peak performance
as can also be seen in the roofline plot in Figure~\ref{fig:rooflineH}.

\subsection{Uniform \mcH-Matrices}

For \mcUH-matrices the above algorithm can be applied directly, only the multiplication of low-rank blocks need
to be modified from \(U_{\clt,\cls} V_{\clt,\cls}^H x|_{\cls}\) to \(\mcW_{\clt} S_{\clt,\cls} \mcX_{\cls}^H
x|_{\cls}\). However, this would be inefficient as the product \(\mcX_{\cls}^H x|_{\cls}\) with the column cluster basis
is performed multiple times as it needs to be computed for each low-rank block in the same block column.
In an analogue way, the product with the row cluster basis \(\mcW_{\clt}\) is performed for each low-rank block in \(\mcM^r_{\clt}\).

Instead, the multiplication with the cluster basis should be performed only once, yielding coefficient vectors
\begin{displaymath}
  s_{\cls} := \mcX_{\cls}^H x|_{\cls}
\end{displaymath}
of size \(k\) which can then be reused for each block in \(\mcM^{c,lr}_{\cls}\), i.e.,
the set of low-rank blocks in \(\mcM_{\cls}\). This process is named \emph{forward transformation} and shown in
Algorithm~\ref{alg:forwarduni}.
\begin{algorithm}{Forward Transformation for \mcUH-matrices}{alg:forwarduni}
  \Procedure{Forward}{$\cls,x$}
  \State \(s_{\cls} := \mcX_{\cls}^H x|_{\cls}\);
  \ForAll{ \(\cls' \in \mcS(\cls)\) }
  \State \function{Forward}[\(\cls',x\)];
  \EndFor
  \EndProcedure
\end{algorithm}

Now let
\begin{equation} \label{eq:sumcoup}
  t_{\clt} := \sum_{\clt,\cls \in \mcM^{r,lr}_{\clt}} S_{\clt,\cls} s_{\cls}
\end{equation}
be the accumulated update by all coupling matrices in \(\mcM^{r,lr}_{\clt}\). The update to \(y\) can then be performed in a single step:
\begin{displaymath}
  y_{\clt} := y_{\clt} + \mcW_{\clt} t_{\clt} .
\end{displaymath}
Thereby also the multiplication with the row cluster bases \(\mcW_{\clt}\) is performed only once, a process also known
as \emph{backward transformation}.

For the parallel evaluation of the \mcUH-matrix-vector product, the same problems as for the
\mcH-matrix case have to be solved, i.e., handling of data races when simultaneously updating identical data blocks.
In case of \mcUH-matrices, the forward transformation is trivially parallelized over all nodes of the cluster
bases tree as all work is on independent data.

This does not hold for the sum in \eqref{eq:sumcoup}. As before for \mcH-matrices, different strategies for handling
this are possible, e.g., guarding updates to \(t_{\clt}\) by a mutex, computing \eqref{eq:sumcoup} in a single
task or as a recursive reduction. Again, a mutex would increase the set of parallel tasks to the set of all low-rank blocks in
\(\mcM^{r,lr}_{\clt}\) whereas the latter approach eliminates collisions at the cost of a smaller parallel
degree. However, the number of nodes in a typical cluster bases tree should be sufficiently large to ensure no idle
times for many-core processors.

Finally, the backward transformation requires further synchronisation between different tasks as overlapping sections of
\(y\) may be updated simultaneously. As for \mcH-matrices, this may be solved by mutex guarded updates (in chunks) or by
preventing collisions by following the hierarchy from the root to the leaves as in Algorithm~\ref{alg:phmvm}, i.e.,
compute the update to \(y_{\clt}\) before any updates to \(y_{\clt'}\) with \(\clt' \in \mcS(\clt)\). This can also be
combined with the computation of \eqref{eq:sumcoup}, which is done in Algorithm~\ref{alg:mvmuni}. There, the set of
dense matrix blocks in a given block row is denoted \(\mcM^{r,d}_{\clt} := \mcM^r_{\clt} \setminus \mcM^{r,lr}_{\clt}\).

\begin{algorithm}{\mcUH-matrix-vector multiplication (without forward trans.)}{alg:mvmuni}
  \Procedure{UniMVM}{$\clt,\mcM^r,y,x$}
  \ForAll{\((\clt,\cls) \in \mcM^{r,lr}_{\clt}\)}
  \State \(t_{\clt} := t_{\clt} + S_{\clt,\cls} s_{\cls}\);
  \EndFor
  \State \(y' := \mcW_{\clt} t_{\clt}\);
  \ForAll{\((\clt,\cls) \in \mcM^{r,d}_{\clt}\)}
  \State \(y' := y' + M_{\clt,\cls} x|_{\cls}\);
  \EndFor
  \State \(y|_{\clt} := y|_{\clt} + y'\)

  \ParForAll{ \(\cls' \in \mcS(\cls)\) }
  \State \function{UniMVM}[\(\cls',\mcM^r,y,x\)];
  \EndFor
  \EndProcedure
\end{algorithm}

In \cite{BruHuyMee:2025} a different strategy was employed which makes use of a slightly different representation of the
coupling matrices \(S_{\clt,\cls}\). Instead of having a single coupling matrix, separate row and column coupling
matrices \(S^{r}_{\clt,\cls}\) and \(S^{c}_{\clt,\cls}\) with \(S_{\clt,\cls} = S^{r}_{\clt,\cls} \cdot
(S^{c}_{\clt,\cls})^H\) are used. These matrices are typically obtained during the cluster basis construction for the
row and the column cluster bases and are normally combined into a single coupling matrix. Depending on the
application the separate storage may lead to a slightly smaller memory footprint for the matrix storage as the rank
of the shared cluster basis is typically larger compared to the original rank of a single block.

Furthermore, one can now decouple the product with \(S_{\clt,\cls}\) into two stages. In \cite{BruHuyMee:2025} the first
stage consists of computing \((S^{c}_{\clt,\cls})^H s_{\cls}\) for each low-rank block, which can be done independent
from all other blocks and also together with the forward transformation. The second stage consists of performing the
product with \(S^{r}_{\clt,\cls}\) followed by the backward transformation and the update of \(y\). The latter is
implemented using thread local data and afterwards combined into the full result.

In all cases the runtime of the \mcUH-matrix-vector multiplication is in \(\landau{n \log n}\) (see
\cite{BruHuyMee:2025}), especially since the cluster bases storage is in \(\landau{n \log n}\).

The results of an approach using mutexes for thread synchronization for the per-block products (``mutex''), of
Algorithm~\ref{alg:mvmuni} (``row wise'') and of the version from \cite{BruHuyMee:2025} (``sep. coupling'') are shown in
Figure~\ref{fig:mvmtime} (center). In all presented cases Algorithm~\ref{alg:mvmuni} yielded the best performance,
followed by the mutex based version and the algorithm using separate coupling matrices from \cite{BruHuyMee:2025}, the
latter again suffering from the additional overhead of combining thread-local results. What is also noteworthy is the
performance advantage compared to \mcH-matrix-vector multiplication. For a fixed accuracy (Figure~\ref{fig:mvmtime},
center, top), also the dependence on the problem size is less than for \mcH-matrices, demonstrating the more efficient
storage scheme of uniform \mcH-matrices.

On the test system the maximal performance of Algorithm~\ref{alg:mvmuni} is 78\% of the peak performance, as can
be seen in Figure~\ref{fig:rooflineH} and as such is almost identical with the performance for \mcH-matrices and again
limited by the memory subsystem.

\subsection{\mcHH-Matrices}

The three steps forward transformation, computation of \(t_{\clt}\) and backward transformation are also present
in the \mcHH-matrix-vector multiplication. However, while for \mcUH-matrices the multiplications with
\(\mcX_{\cls}\) and \(\mcW_{\clt}\) are standard matrix-vector multiplications, the corresponding operations for \mcHH
cluster bases are indirect due to its recursive definitions. For the first step we have
\begin{displaymath}
  s_{\cls} := \mcX_{\cls}^H x|_{\cls} = \sum_{\cls' \in \mcS(\cls)} E_{\cls'}^H \left( \mcX_{\cls'}^H x|_{\cls'} \right )
\end{displaymath}
which leads to a recursive evaluation of the child cluster bases. Only at the leaves, a direct computation is performed
and the resulting coefficients are handed back to the parent cluster bases for updating the corresponding local
coefficients via the transfer matrices. This process is performed until the root of the cluster basis tree is reached as
shown in Algorithm~\ref{alg:forwardh2}.

\begin{algorithm}{Forward Transformation for \mcHH-matrices}{alg:forwardh2}
  \Procedure{Forward}{$\cls,x$}
  \If{\(\cls \in \mcL(T_J)\)}
  \State \(s_{\cls} := \mcX_{\cls}^H x|_{\cls}\);
  \Else
  \ForAll{ \(\cls' \in \mcS(\cls)\) }
  \State \function{Forward}[\(\cls',x\)];
  \State \(s_{\cls} := s_{\cls} + E_{\cls'}^H s_{\cls'}\);
  \EndFor
  \EndIf
  \EndProcedure
\end{algorithm}

\begin{remark}
  Though Algorithm~\ref{alg:forwardh2} is very simular to Algorithm~\ref{alg:forwarduni}, the latter can compute all
  coefficient vectors in parallel while a strict leaves-to-root dependency exists for the \mcHH-matrix version.
\end{remark}

In an analogue way, the backward transformation with the row cluster basis \(\mcW_{\clt}\) is evaluated in reverse order,
i.e., starting at the root and shifting locally transformed coefficients to child cluster bases until the leaves are
reached where the actual update of the destination vector \(y\) can be computed. We refer the reader to
\cite{Boerm:2010} for a detailed description of the full process.

A cluster basis centric approach to parallelization as is used for \mcUH-matrices can also be found in the literature
for \mcHH-matrices, e.g., in \cite{BoeBen:2008} for the distributed memory case or in \cite{BouTurKey:2019} for GPUs. We
adopt Algorithm~\ref{alg:mvmuni} for \mcHH-matrices, i.e., combine local coupling matrix computations and the backward
transformation by following the hierarchy of the row cluster basis. The only change for \mcHH-matrices is the backward
transformation which either restricts local coefficients to child clusters or applies the final result to the
destination vector. Since updates to \(y\) are only performed at the leaf clusters also no race condition exists between
simultaneously executed tasks on different processor cores. The result is presented in Algorithm~\ref{alg:mvmh2}.

\begin{algorithm}{\mcHH-matrix-vector multiplication (without forward trans.)}{alg:mvmh2}
  \Procedure{H2MVM}{$\clt,\mcM^r,y,x$}
  \ForAll{\((\clt,\cls) \in \mcM^{r,lr}_{\clt}\)}
  \State \(t_{\clt} := t_{\clt} + S_{\clt,\cls} s_{\cls}\);
  \EndFor
  \If{\(\clt \in \mcL(T_I)\)}
  \State \(y' := \mcW_{\clt} t_{\clt}\);
  \Else
  \ForAll{\(\clt' \in \mcS(\clt)\)}
  \State \(t_{\clt'} := E_{\clt'} t_{\clt}\);
  \EndFor
  \EndIf
  \ForAll{\((\clt,\cls) \in \mcM^{r,d}_{\clt}\)}
  \State \(y' := y' + M_{\clt,\cls} x|_{\cls}\);
  \EndFor
  \State \(y|_{\clt} := y|_{\clt} + y'\)

  \ParForAll{ \(\cls' \in \mcS(\cls)\) }
  \State \function{H2MVM}[\(\cls',\mcM^r,y,x\)];
  \EndFor
  \EndProcedure
\end{algorithm}

The main advantage of matrix-vector multiplication in the \mcHH-matrix format compared to \mcH- and \mcUH-matrices is
the runtime of \(\landau{n}\) (see \cite{Boerm:2010}).

As for \mcUH-matrices also a mutex based matrix-vector multiplication was implemented, i.e., the updates
in \eqref{eq:sumcoup} to \(t_{\clt}\) are guarded by a mutex associated to \(t_{\clt}\).

Figure~\ref{fig:mvmtime} (right) shows the results for both versions. The runtime for the \mcHH-matrix-vector
multiplication is very similar to the corresponding runtime of the \mcUH-matrix version. However, the linear runtime and
storage complexity of \mcHH-matrices result in a slight advantage compared to \mcUH-matrices, which should grow for even
larger problem sizes. Furthermore, as for \mcUH-matrices a mutex based approach results in higher runtimes compared to the
synchronization free version in Algorithm~\ref{alg:mvmh2}.

As is shown in Figure~\ref{fig:rooflineH} the maximal performance of Algorithm~\ref{alg:mvmh2} is about 82\% of the peak
performance of the test system. This is slightly better compared to the \mcUH-matrix-vector multiplication and owed to
the fact that the forward and backward transformations require significantly less data, thereby reducing the stress on
the memory subsystem as this is again a memory bandwidth limited computation.


\section{Compressed \mcH-Matrices} \label{sec:zhmat}

\subsection{Floating Point Compression} \label{sec:zhmatflt}

Floating point data in hierarchical matrices appears in inadmissible blocks as dense matrices holding the coefficients and in
low-rank blocks in the form of the low-rank factors or as coupling matrices in \mcUH- or \mcHH-matrices. Often these are
stored in the FP64 (or FP32) format. However, due to low-rank approximation with accuracy \(\varepsilon\), already an error
is introduced which is typically much larger than the unit roundoff of FP64 (or even FP32).

In \cite{KriLtaLuo:2022,Kri:2025} the FP64 storage was replaced by error adaptive floating point compression, i.e., an
optimized storage format was chosen with a representation error depending on \(\varepsilon\). Different compressors are
available to implement such a direct compression of floating point data, e.g., ZFP \cite{Lindstrom:2014} or BLOSC
\cite{blosc}. Furthermore, different storage schemes based on the IEEE-754 floating point standard were examined.

For one such format (AFLP in \cite{Kri:2025}) the number of mantissa bits \(m_{\varepsilon}\) is chosen based on the
low-rank approximation error \(\varepsilon\) as
\begin{displaymath}
  m_{\varepsilon} := \left\lceil -\log_2 \varepsilon \right\rceil.
\end{displaymath}
Similarly, the number of exponent bits \(e_{\operatorname{dr}}\) is reduced to meet the \emph{dynamic range} of the
data, i.e., the logarithm of the ratio between the largest (\(v_{\max}\)) and smallest (\(v_{min}\)) absolute value:
\begin{displaymath}
  e_{\operatorname{dr}} := \left\lceil \log_2 \log_2 \frac{v_{\max}}{v_{\min}} \right\rceil.
\end{displaymath}
The data values are scaled and shifted such that the exponent (including bias) is at least one. This way the lowest
\(e_{\operatorname{dr}}\) bits hold the exponent and can easily be extracted\footnote{Also the highest exponent bit is
  always 1.}.

While storing just \(1 + m_{\varepsilon} + e_{\operatorname{dr}}\) bits per value is space efficient, reading from or
writing to memory requires costly bit manipulations. Instead, for AFLP \(m_{\varepsilon}\) is increased to
\(m'_{\varepsilon}\) such that \(1 + m'_{\varepsilon} + e_{\operatorname{dr}}\) is a multiple of 8 which makes memory
operations byte aligned.

\begin{figure}[htb]
  \centering
  \pgfdeclareimage[width=.25\textwidth]{fpx}{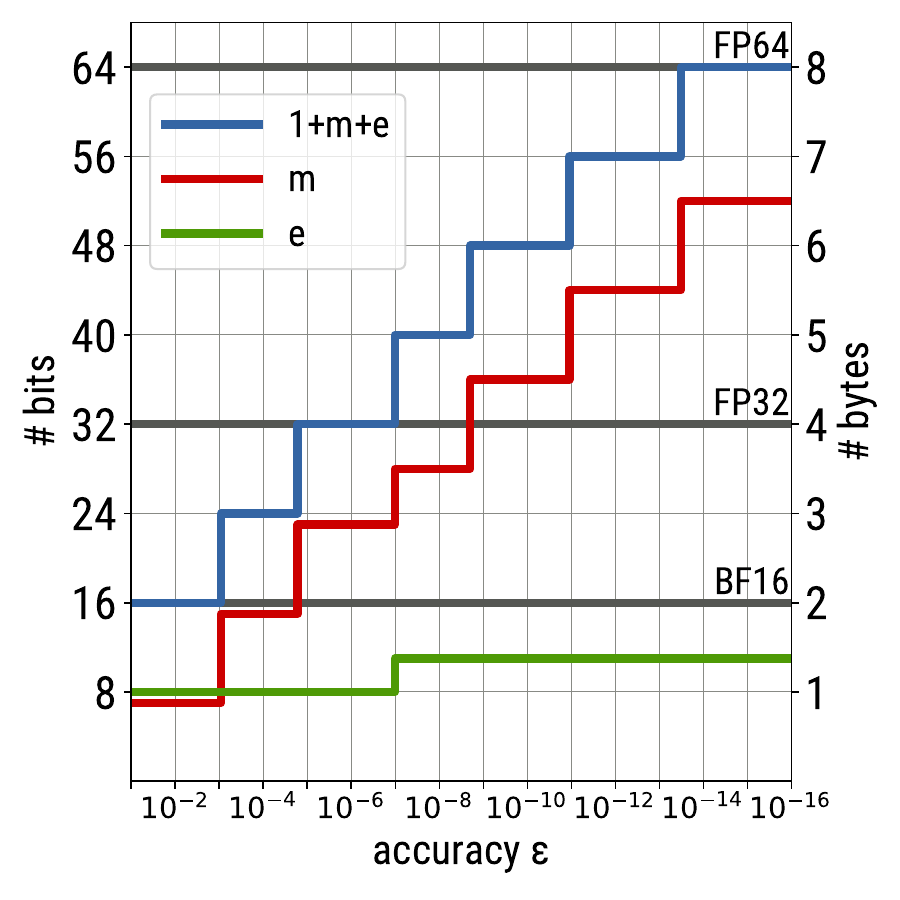}
  \raisebox{2.3cm}{\parbox{.23\textwidth}{%
  \begin{tabular}{rrr}
    & \multicolumn{2}{c}{Format} \\
    \multicolumn{1}{c}{\(\lceil -\log_2 \varepsilon \rceil\)}
    & \multicolumn{1}{c}{\(e\)}
    & \multicolumn{1}{c}{\(m\)} \\
    \toprule
    \( 0 \ldots 10\) &  8 &  7 \\
    \(11 \ldots 15\) &  8 & 15 \\
    \(16 \ldots 23\) &  8 & 23 \\
    \(24 \ldots 28\) & 11 & 28 \\
    \(29 \ldots 36\) & 11 & 36 \\
    \(37 \ldots 44\) & 11 & 44 \\
    \(45 \ldots 52\) & 11 & 52 \\
  \end{tabular}}}\hspace{-3mm}\pgfuseimage{fpx}
  \caption{Custom floating point formats based on accuracy}
  \label{tab:fpx}
\end{figure}

Also discussed in \cite{Kri:2025} are formats with an adaptive mantissa length but a fixed exponent size of 8 or 11 bits
as defined by the FP32/FP64 formats (named SFL and DFL). In \cite{AmeJegExcMarPic:2025} this is combined into a single
format. Depending on the accuracy the number \(m\) of mantissa bits and \(e\) exponent bits are chosen such that the resulting
format is a byte-aligned, truncated version of the standard FP32 or FP64 formats as is shown in Figure~\ref{tab:fpx}. As
described in \cite{AmeJegExcMarPic:2025} the major advantage of such a scheme is, that conversion to and from FP64 (or
FP32) can be done by using fast AVX512 SIMD instructions. In the following we will call this scheme \emph{FPX}
(\emph{F}loating \emph{P}oint e\emph{X}tended)\footnote{In \cite{AmeJegExcMarPic:2025} a common name is missing as each
 individual (sub) format was considered separately.}. However, in constrast to the rounding strategy employed in
\cite{AmeJegExcMarPic:2025}, where the most significant bit of the truncature is set to 1, round-to-nearest (RTN) is
used in this work.

\begin{remark} \label{rem:fpxspeed}
  Since only byte shifting is needed for decompression with FPX, this compression format is up to 50\% faster on the
  test hardware compared to AFLP, which requires floating point addition and multiplication.
\end{remark}

\begin{figure}[htb]
  \centering
  \pgfdeclareimage[width=.31\textwidth]{zerror}{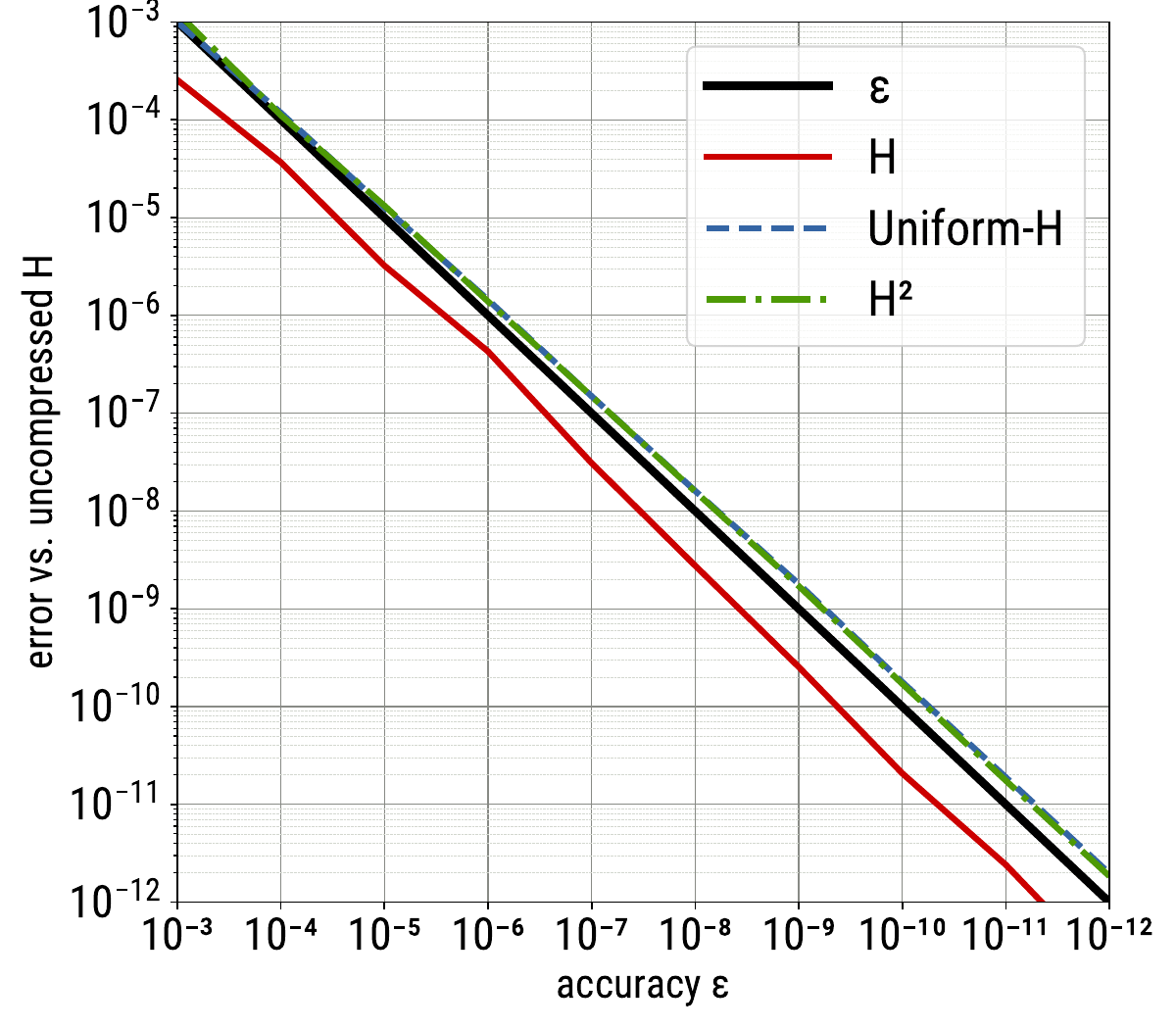}
  \pgfuseimage{zerror}
  \caption{Error of AFLP compressed \mcH, \mcUH and \mcHH-matrices compared to reference uncompressed \mcH-matrix.}
  \label{fig:zerror}
\end{figure}

Independent on the particular choice of the compression scheme (AFLP or FPX), this so called \emph{direct} compression mode is applied to
the dense data of inadmissible blocks. Furthermore, for \mcUH-matrices and \mcHH-matrices the coupling matrices for each
low-rank block are also compressed based on direct compression. This also holds for the transfer matrices of
\mcHH-matrix cluster bases. While for the inadmissible blocks this introduces an additional error of \(\varepsilon\),
this does not affect the overall error of the \mcH, \mcUH or \mcHH-matrix as already low-rank approximation was applied
with an approximation error of \(\varepsilon\) per low-rank block. For the same reason does the compression of the
coupling matrices in \mcUH and \mcHH-matrices not increase the overall error. This can be seen in
Figure~\ref{fig:zerror}, where the error of the compressed matrices vs. an uncompressed (reference) \mcH-matrix is
shown for the AFLP compression. All formats closely follow the predefined approximation error \(\varepsilon\).

\subsection{Low-Rank Matrices} \label{sec:zhmatlr}

As introduced in \cite{AmeBoiBut:2022} and combined with general floating point compression schemes in
\cite{Kri:2025}, low-rank matrices permit an advanced compression scheme with a \emph{V}ariable \emph{A}ccuracy for each
column of the \emph{L}ow-\emph{R}ank factors (\emph{VALR}). For a block \(M_{\clt,\cls}\) we assume a rank-\(k\)
approximation \(U_{\clt,\cls}\cdot V_{\clt,\cls}^H\) with \(\|M_{\clt,\cls} - U_{\clt,\cls} V_{\clt,\cls}^H \|_F \le \delta\), e.g., \(\delta = \varepsilon /
\|M_{\clt,\cls}\|_F\). Using the singular value decomposition we can find orthogonal matrices \(W_{\clt,\cls}\) and \(X_{\clt,\cls}\) and a
diagonal matrix \(\Sigma_{\clt,\cls} = \operatorname{diag}(\sigma_0,\ldots,\sigma_{k-1})\) with the singular values \(\sigma_0 >
\sigma_1 > \ldots \sigma_{k-1}\) of \(U_{\clt,\cls} V_{\clt,\cls}^H\) such that \(U_{\clt,\cls} V_{\clt,\cls}^H = W_{\clt,\cls} \Sigma_{\clt,\cls} X_{\clt,\cls}^H\).

If the \(i\)'th column \(w_i\) of \(W_{\clt,\cls}\) and \(x_i\) of \(X_{\clt,\cls}\) are stored with accuracy \(\delta_i
:= \delta/\sigma_i\), resulting in vectors \(\tilde w_i,\tilde x_i\) and corresponding matrices \(\widetilde
W_{\clt,\cls}\) and \(\widetilde X_{\clt,\cls}\), then the total approximation error is (see \cite[Section~4]{Kri:2025})
\begin{equation} \label{eqn:valr}
  \|M_{\clt,\cls} - \widetilde W_{\clt,\cls} \Sigma_{\clt,\cls} \widetilde X_{\clt,\cls}^H \|_F
  \le
  \delta \left( 1 + 2 k + \delta \sum_{i=1}^k \frac{1}{\sigma_i} \right)
\end{equation}
With this, any direct floating point compression method can be used to yield an improved storage method for low-rank
matrices. The main advantage of this scheme compared to direct compression is, that in the latter case the chosen
\emph{fixed} precision is applied to the full data whereas with VALR even for a high accuracy, a low precision may be
used for some part of the data.

The cluster bases \(\mcW_{\clt} \in \R^{\#\clt \times k}\) in \mcUH or \mcHH-matrices are often computed using the
singular value decomposition of some intermediate data, e.g., when combining block-local cluster bases
(rf. \cite{Boerm:2010,BruHuyMee:2025}). With this one typically has as a final step \(W_{\clt} = \mcW_{\clt}
\Sigma_{\clt} X_{\clt}\) for some intermediate matrix \(W_{\clt}\) with the singular values again in \(\Sigma_{\clt}\)
normally being discarded and only \(\mcW_{\clt}\) being used. However, the singular values allow the application of VALR
compression to the cluster bases. For this let \(\mcW_{\clt} = \set{w_0,\ldots,w_{k-1}}\). Furthermore, let \(\widetilde
\mcW_{\clt} := \set{\widetilde w_0, \ldots, \widetilde w_{k-1}}\) be a compressed version of \(\mcW_{\clt}\) with
\(\|w_i - \widetilde w_i\|_2 \le \delta_i\), \(0 \le i < k\). Then the total error for the compressed cluster basis is
\begin{equation} \label{eqn:valrcb}
  \|\mcW_{\clt} \Sigma_{\clt} - \widetilde \mcW_{\clt} \Sigma_{\clt}\|_F
  \le \sum_{i=0}^{k-1} \delta_i \sigma_i
  = k \delta .
\end{equation}

When applying the local error per \(\sigma_i\) the additional factor \(k\) in \eqref{eqn:valrcb} is taken into account
to compensate for the error increase. In a similar way the factor in \eqref{eqn:valr} is used to adjust local error
bounds (rf. \cite[Section 4]{Kri:2025}). The resulting global error stays close to the low-rank approximation error as
is visible in Figure~\ref{fig:zerror}.

\begin{figure}[htb]
  \centering
  \pgfdeclareimage[width=.24\textwidth]{memzdim}{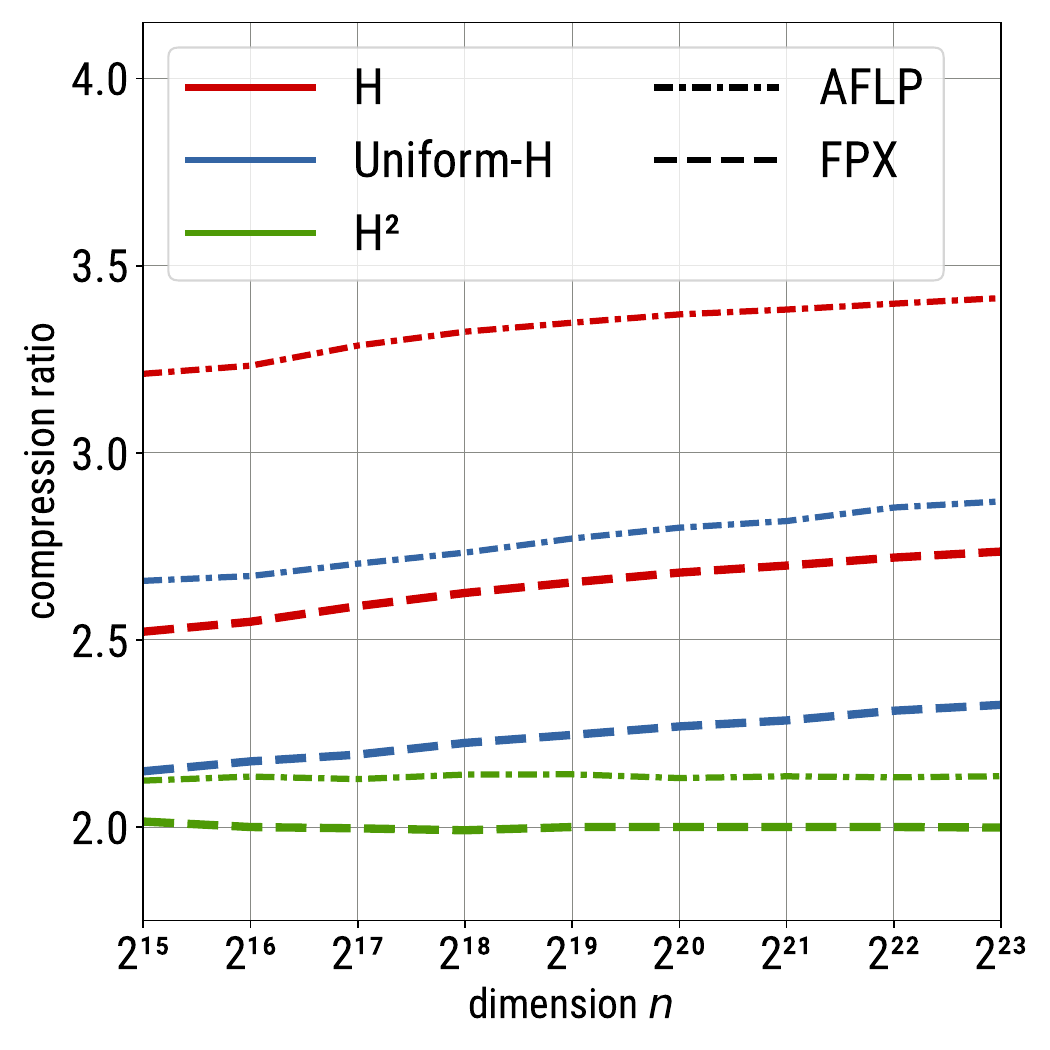}
  \pgfdeclareimage[width=.24\textwidth]{memzeps}{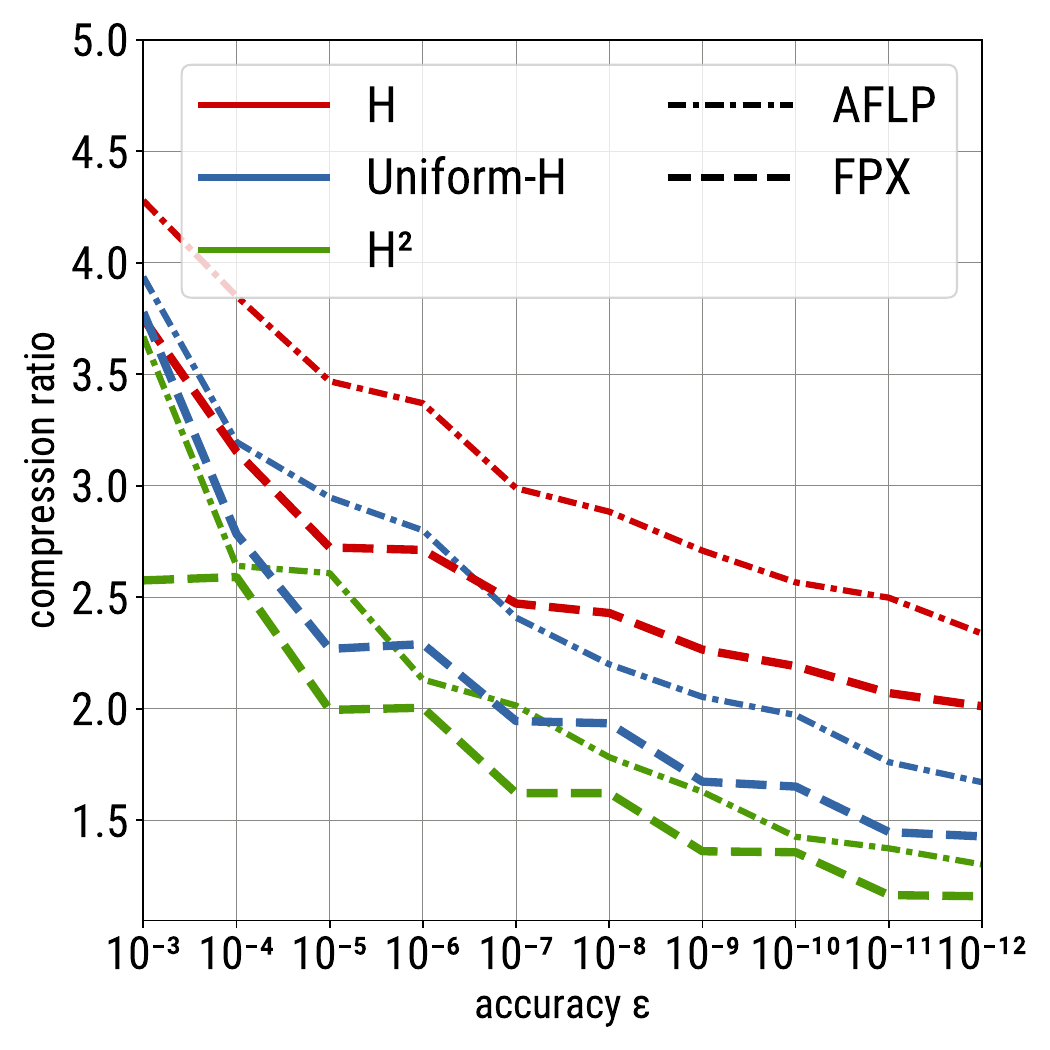}
  \pgfuseimage{memzdim}\pgfuseimage{memzeps}
  \caption{Compression rates of AFLP and FPX depending on matrix size (left) and accuracy (right).}
  \label{fig:comprates}
\end{figure}

As VALR can be applied to every low-rank block in \mcH-matrices, it is expected that best compression rates are achieved
for this matrix format. In the case of \mcUH-matrices one can still apply VALR to all cluster bases, whereas for
\mcHH-matrices only the leaf cluster bases may be compressed using VALR and therefore are expected to show the least
reduction in memory when using compressed storage. This is clearly visible in the results in Figure~\ref{fig:comprates}
where the compression ratios for \mcH-, \mcUH and \mcHH-matrices are shown for
different matrix sizes and accuracies. \mcH-matrices show the best compression rates, followed by \mcUH-matrices. What
is also notable is that the compression rates increase with the matrix size for \mcH- and \mcUH-matrices whereas they
stay constant for \mcHH-matrices. For \mcH-matrices this is due to the increasing portion that low-rank blocks have on
the total memory costs, which grows from 67\% for the smallest problem size to about 82\% for the largest problems.

Finally, in the VALR compression for low-rank blocks, single vectors of the low-rank factors are handled separately
and the values within such a vector are typically of similar magnitude. Therefore, the adaptivity of AFLP for the
exponent size is of particular advantage compared to FPX yielding better compression ratios for all matrix formats.

\begin{figure}[htb]
  \centering
  \pgfdeclareimage[width=.24\textwidth]{memvsdim}{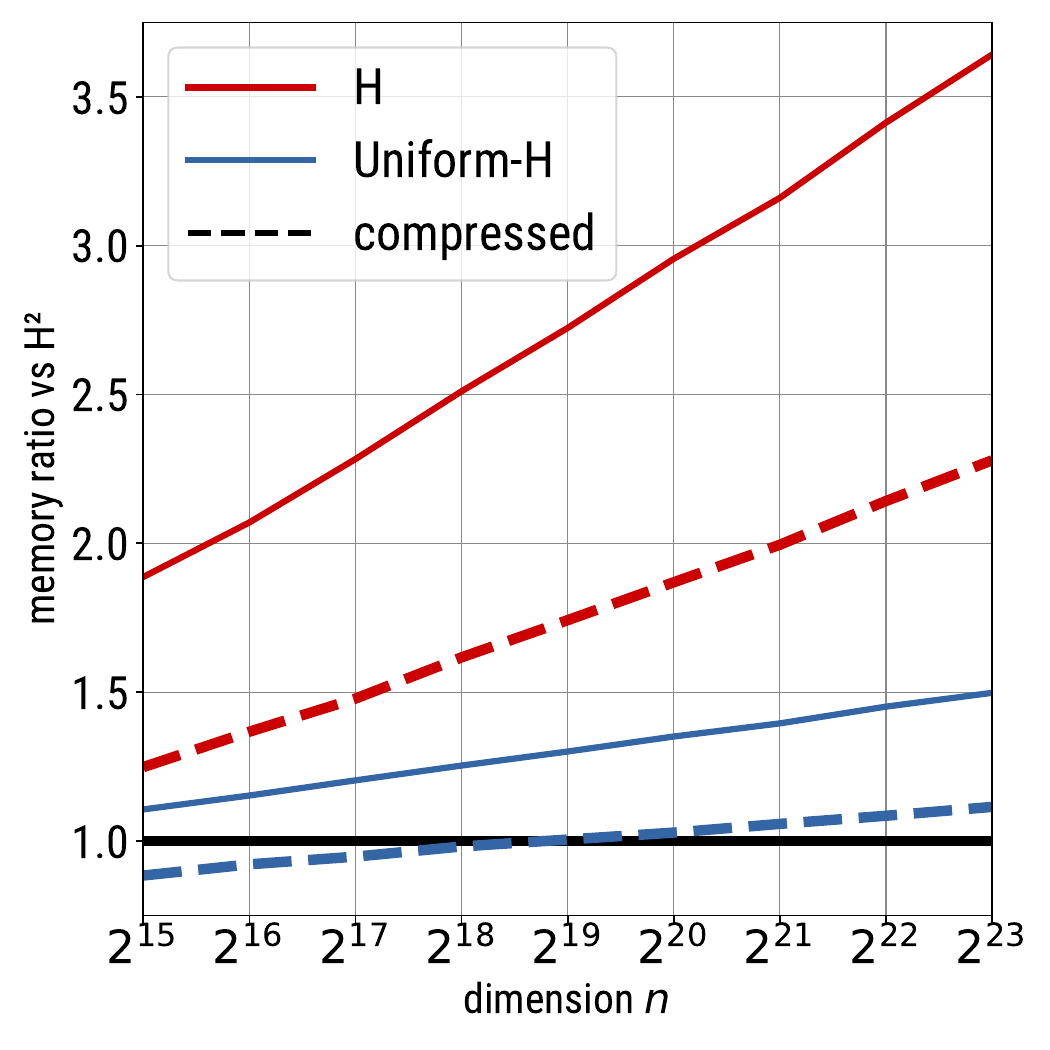}
  \pgfdeclareimage[width=.24\textwidth]{memvseps}{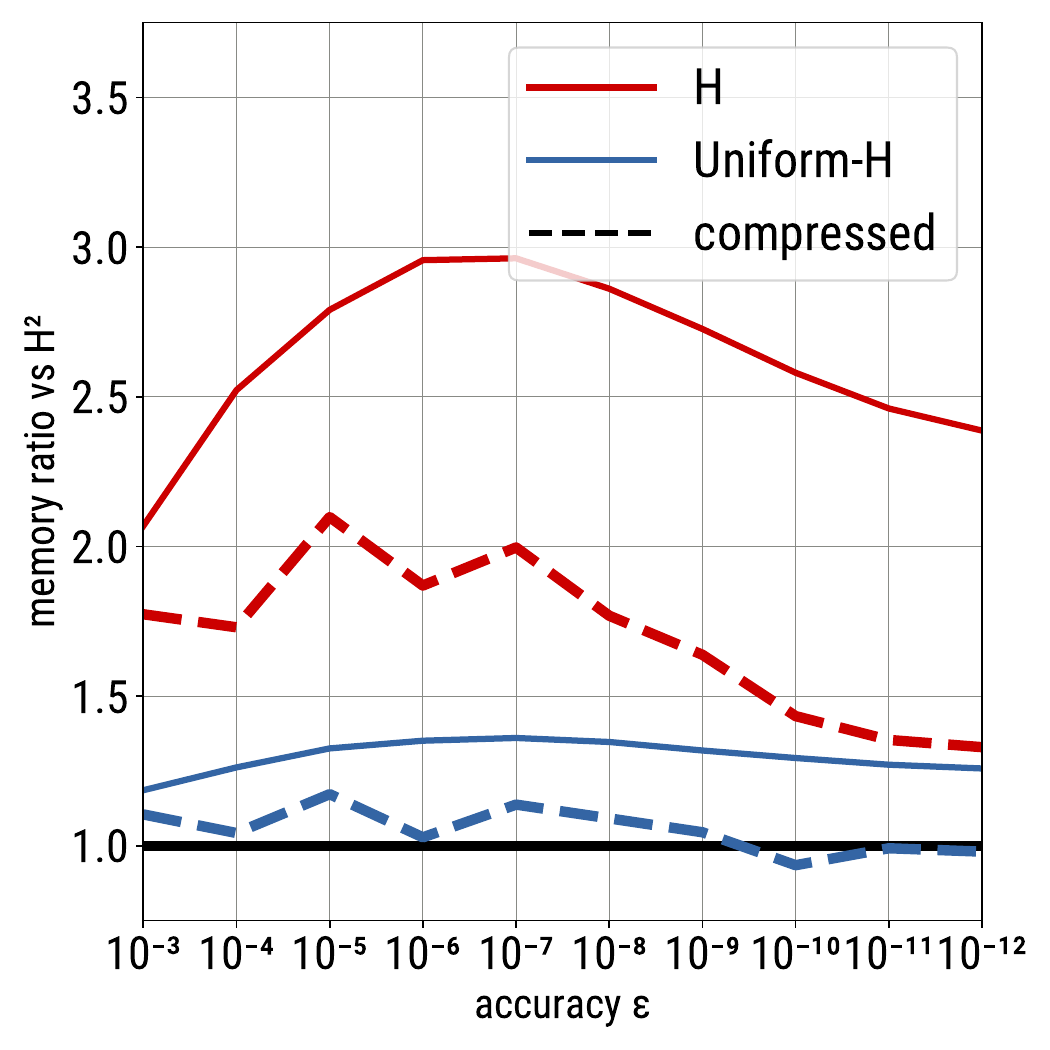}
  \pgfuseimage{memvsdim}\pgfuseimage{memvseps}
  \caption{Comparison of memory of uncompressed and compressed \mcH- and \mcUH-matrix format with \mcHH-format depending on matrix size (left) and accuracy (right).}
  \label{fig:memvsh2}
\end{figure}

Since compression ratios for the \mcH and the \mcUH-format are better compared to the \mcHH-format, the advantage of the
latter is reduced with compression enabled. This can be seen in Figure~\ref{fig:memvsh2}, where the ratio of the memory
size between the \mcH- and the \mcUH-format and the \mcHH-format are shown for uncompressed and compressed
storage. For smaller matrix sizes compressed \mcUH-matrices are even more memory efficient that compressed
\mcHH-matrices. Only the better memory complexity of the \mcHH-format leads to less memory for larger matrix sizes. For
a varying accuracy, \mcUH-matrices are almost identical to \mcHH-matrices for the particular matrix size. While storage
for \mcH-matrices does not come as close to the storage of \mcHH-matrices as \mcUH-matrices do, the advantage of the
\mcHH-format compared to the \mcH-format is significantly reduced.

\begin{figure}[htb]
  \centering
  \pgfdeclareimage[width=.24\textwidth]{hodlrblrmem}{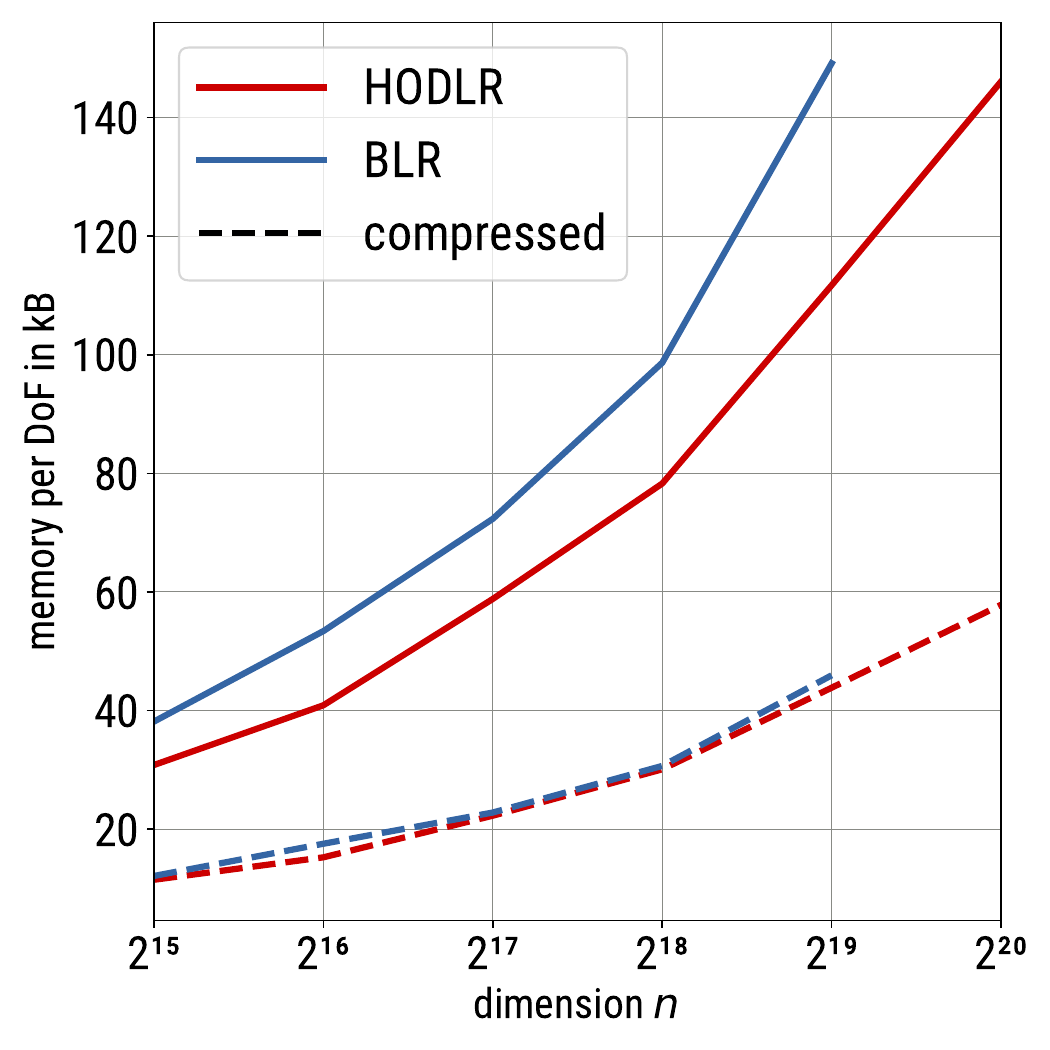}
  \pgfdeclareimage[width=.24\textwidth]{hodlrblrrat}{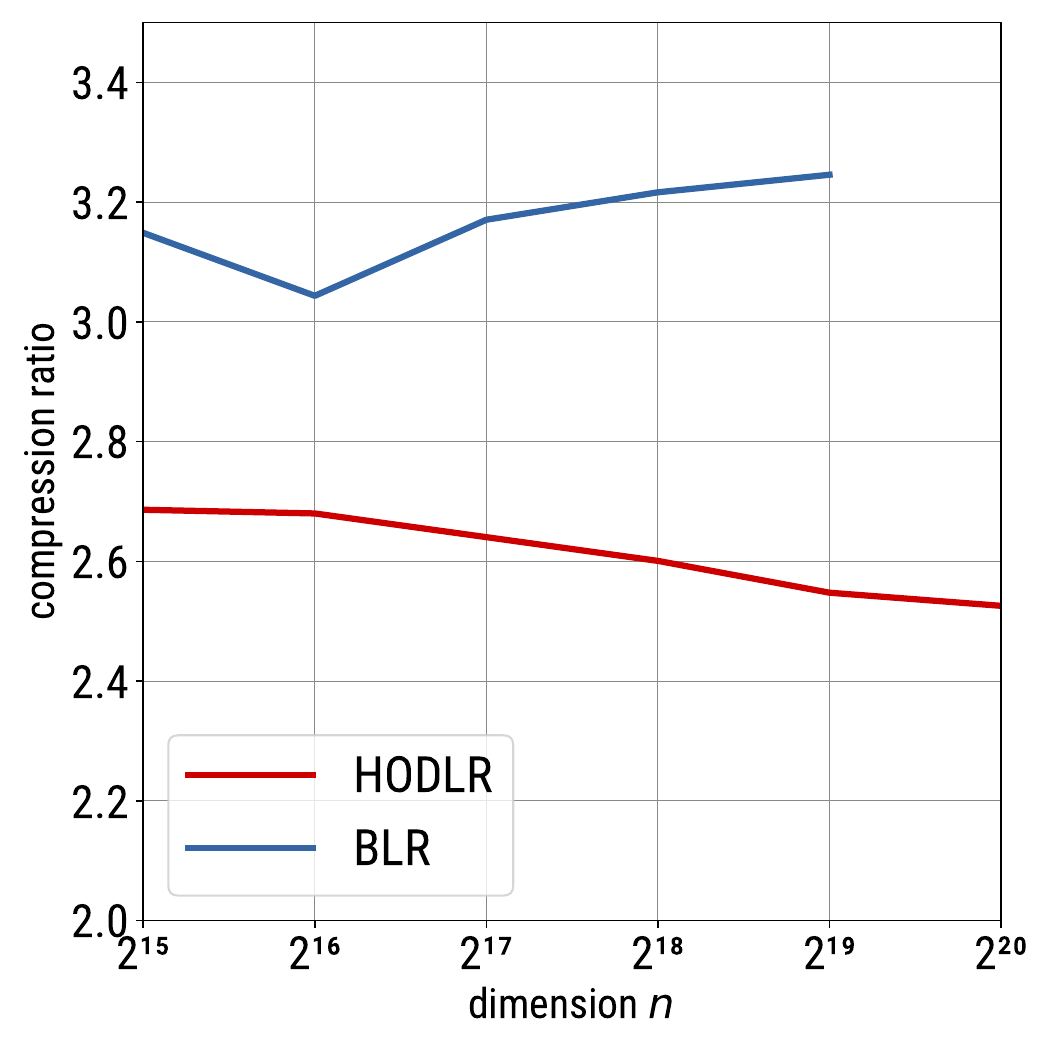}
  \pgfuseimage{hodlrblrmem}\pgfuseimage{hodlrblrrat}
  \caption{Memory of uncompressed and compressed HODLR and BLR matrices (left) and corresponding compression ratios (right).}
  \label{fig:memhodlr}
\end{figure}

Finally, to demonstrate that the technique is not limited to standard \mcH-matrices, Figure~\ref{fig:memhodlr} shows the
memory sizes and compression ratios for the HODLR and the BLR format when representing the same matrix. Of particular
interest is, that the compressed size of both formats is basically identical, even though HODLR is more efficient in the
uncompressed case.

\subsection{Compressed \mcH-Matrix-Vector Multiplication}

The performance of \mcH-MVM when using floating point compression was already a (minor) topic of \cite{Kri:2025}. There,
for a dense or low-rank block, the compressed data was first fully converted from the storage format into the compute
format and only then the local matrix-vector multiplication was performed. This way, the standard arithmetic (BLAS)
kernels, typically optimized by hardware vendors, could be reused. This is more closely investigated in
\cite{AmeJegExcMarPic:2025} by splitting the compressed data into sub-blocks such that the fast memory
close to the processor cores is used in an optimal way.

In \cite{AnzGruQui:2019} the concept of a \emph{memory-accessor} is described which implements on-the-fly conversion
between the storage format and the computation format during the arithmetic. The disadvantage of this approach is, that
one can not use the optimized BLAS routines but needs to implement special functions in the compressed format. However,
as matrix-vector multiplication for \mcH-, \mcUH- and \mcHH-matrices is often memory bandwidth limited it may be more
forgiving for a (potentially) less heavily optimized implementation. Furthermore, in principle only a single function
for dense matrix-vector multiplication is needed in Algorithms~\ref{alg:phmvm}, \ref{alg:mvmuni} and \ref{alg:mvmh2} for
dense and low-rank blocks, reducing the implementation effort.

In any case, such an approach requires fast access to the compressed values, which holds for the above described
compression schemes AFLP and FPX but much less so for ZFP or BLOSC. Furthermore, especially ZFP showed a low
decompression performance compared to other formats (see \cite{Kri:2025}). However, by tightly coupling the compression scheme with
the matrix-vector multiplication, any compression format could be used. As such, the restriction to AFLP and FPX in this
work should be considered a proof-of-concept.

\begin{figure*}[htb]
  \centering
  \pgfdeclareimage[width=.31\textwidth]{zmvmdH}{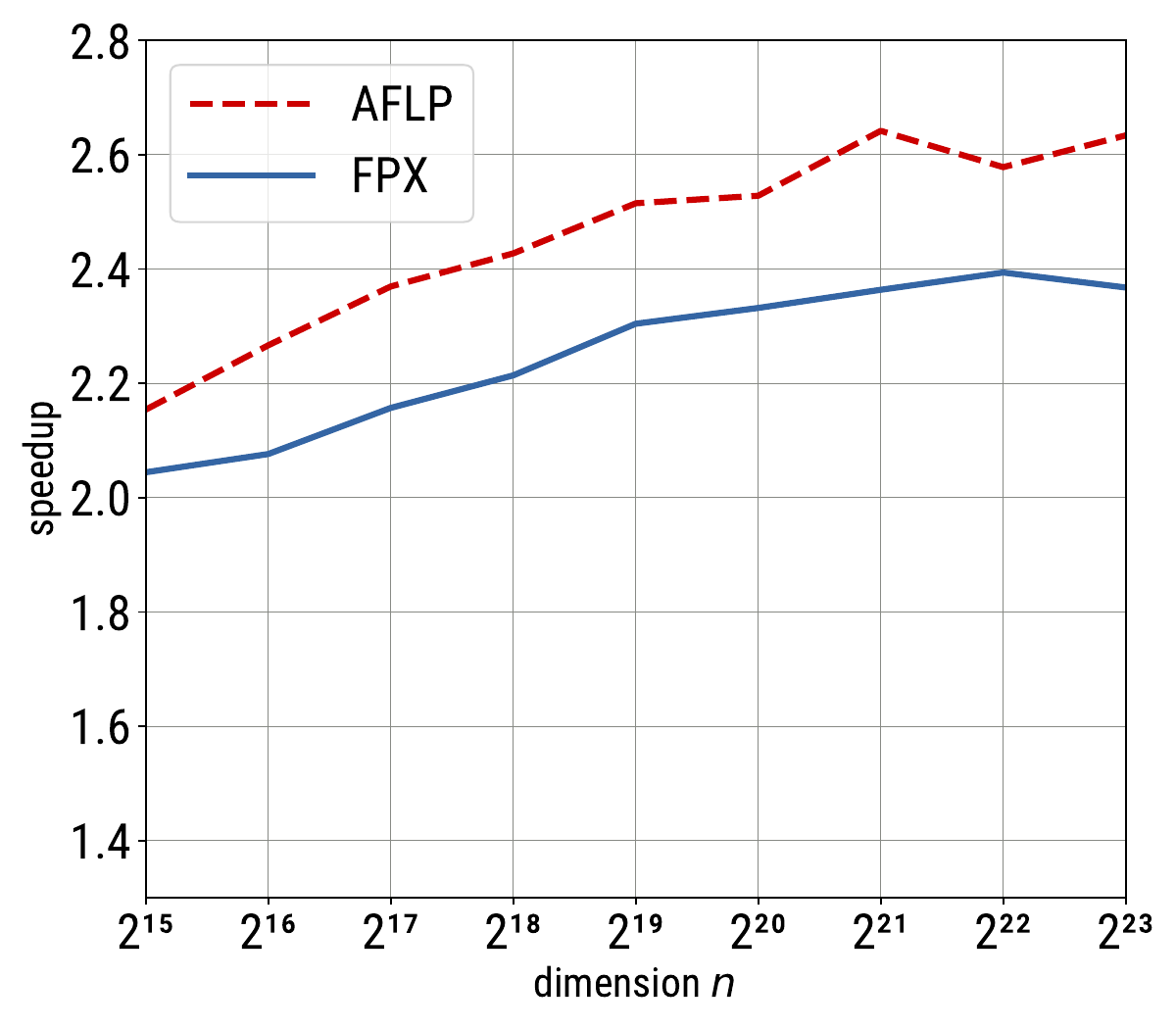}
  \pgfdeclareimage[width=.31\textwidth]{zmvmdU}{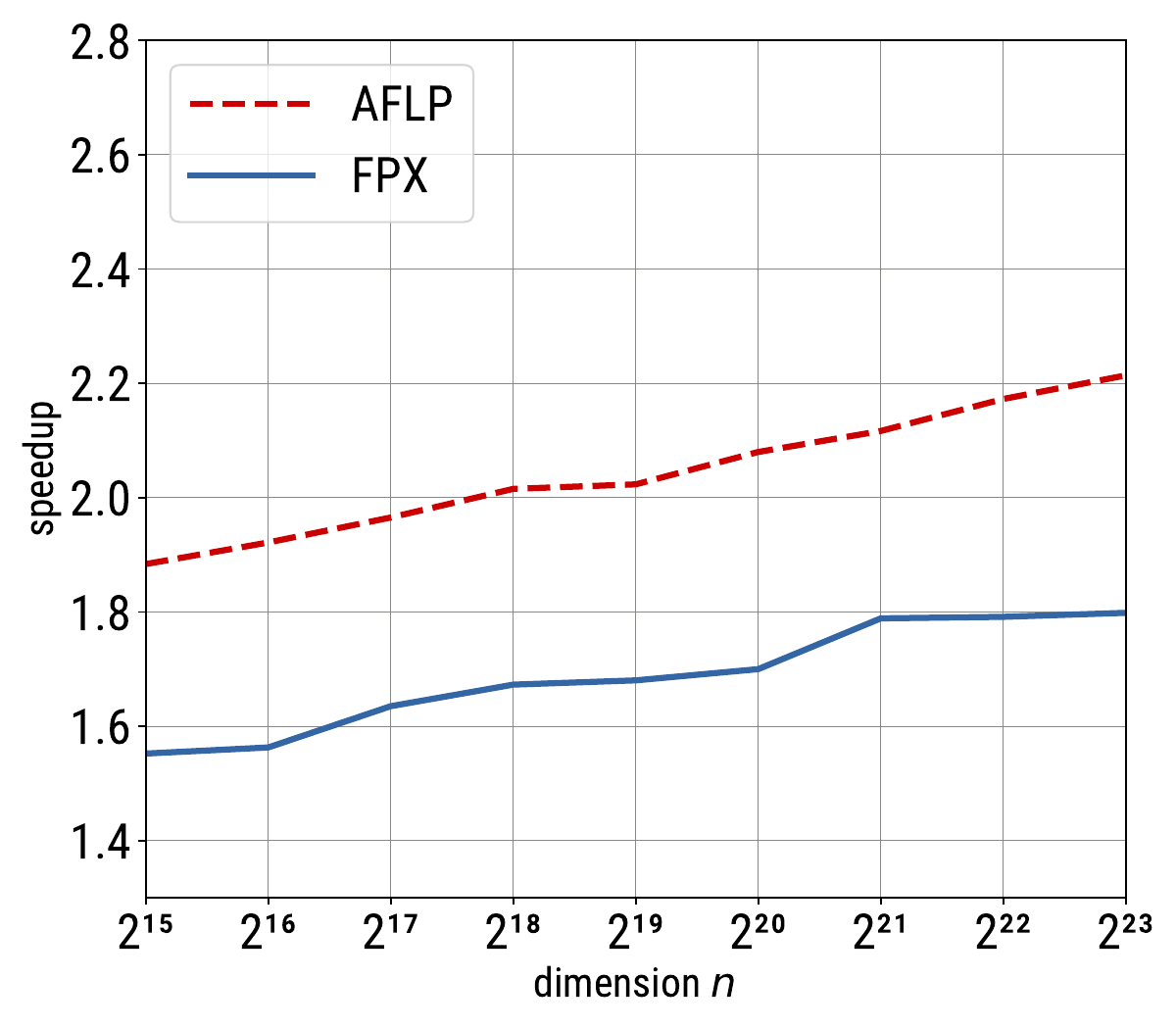}
  \pgfdeclareimage[width=.31\textwidth]{zmvmd2}{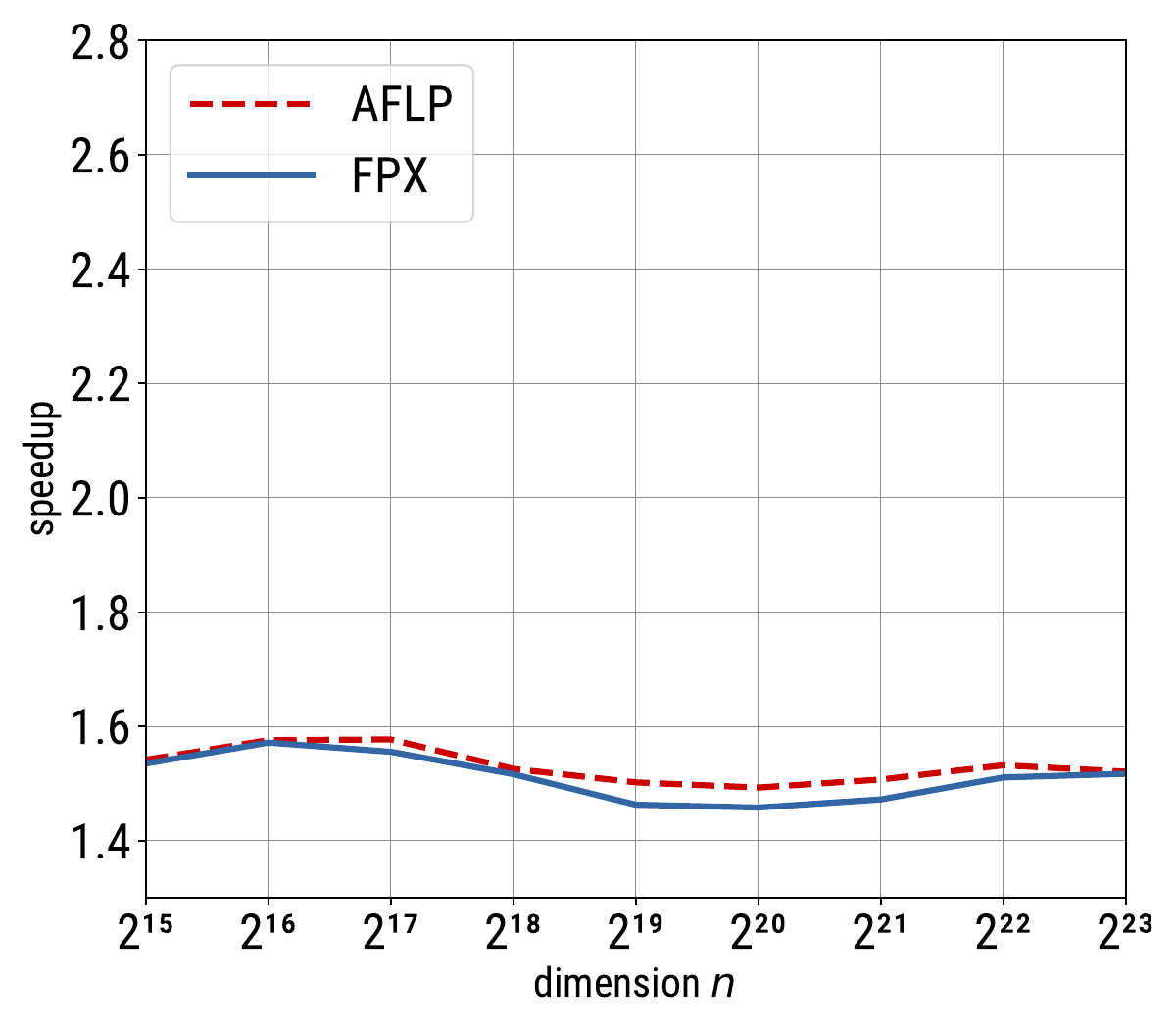}
  \pgfdeclareimage[width=.31\textwidth]{zmvmeH}{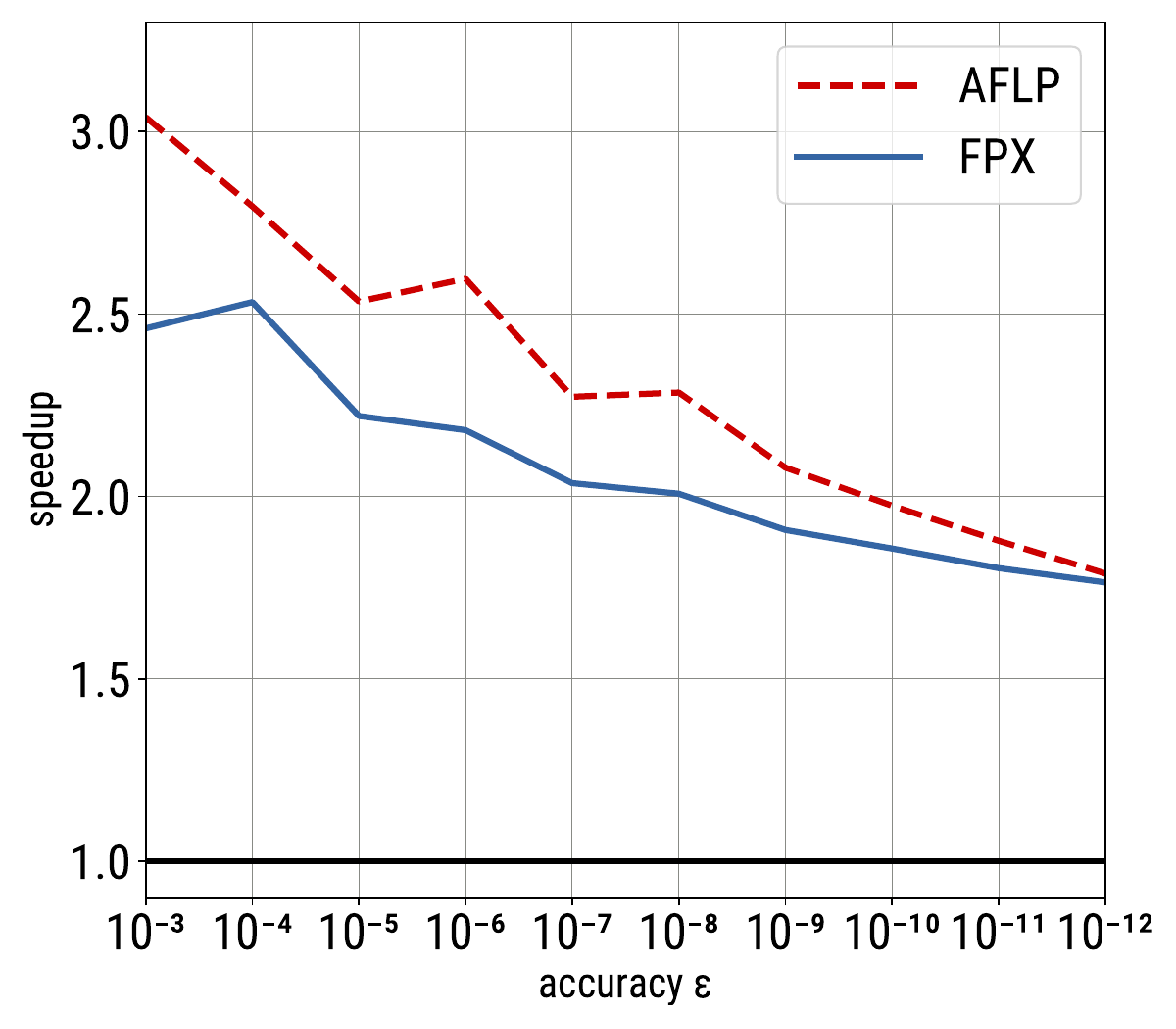}
  \pgfdeclareimage[width=.31\textwidth]{zmvmeU}{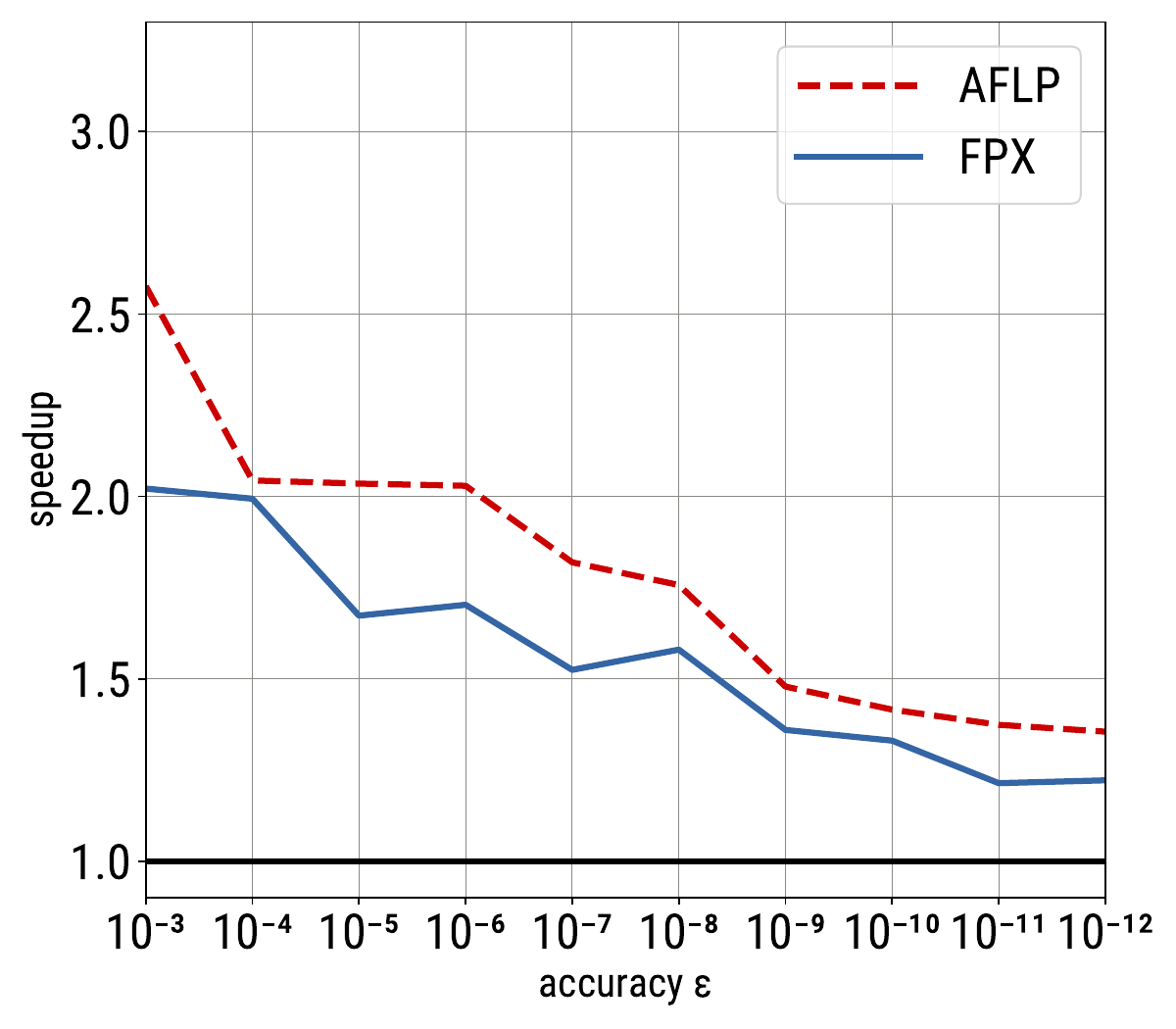}
  \pgfdeclareimage[width=.31\textwidth]{zmvme2}{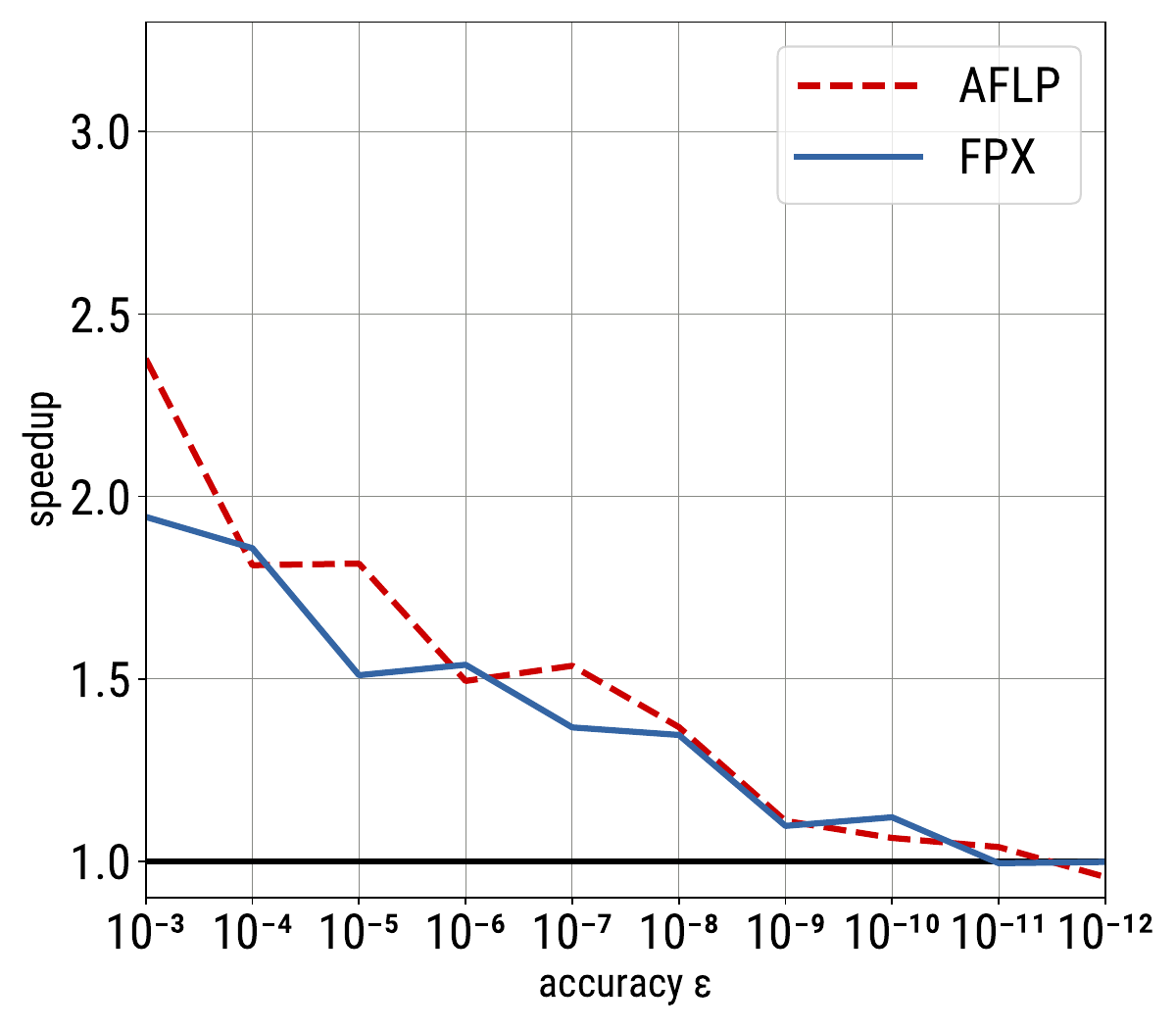}
  \begin{tabular}{ccc}
    \mcH & \mcUH & \mcHH \\
  \pgfuseimage{zmvmdH} & \pgfuseimage{zmvmdU} & \pgfuseimage{zmvmd2}\\
  \pgfuseimage{zmvmeH} & \pgfuseimage{zmvmeU} & \pgfuseimage{zmvme2}
  \end{tabular}
  \caption{Speedup of compressed matrix-vector multiplication for \mcH (left), \mcUH (center) and
    \mcHH-matrices (right) compared to uncompressed multiplication.}
  \label{fig:zmvmtime}
\end{figure*}

The basic implementation of the compressed dense matrix-vector multiplication required in
Algorithms~\ref{alg:phmvm}, \ref{alg:mvmuni} and \ref{alg:mvmh2} is shown in Algorithm~\ref{alg:zdmvm} for the
application of a non-transposed \(n \times m\) matrix \(D\). It is straight-forward, without particular optimizations, aside
from standard code reorganization due to the use of column-major storage scheme. Only the access to the coefficients in
\(D\) is replaced by the corresponding coefficient decompression.

\begin{algorithm}{Matrix-vector multiplication with compressed dense \(n \times m\) matrix}{alg:zdmvm}
  \Procedure{zmvm}{\texttt{in}: $D, x$, \texttt{inout}: $y$}
    \For{ \(0 \le j < m\) }
      \For{ \(0 \le i < n\) }
        \State \(y_i := y_i + \function{decompress}[\(D_{ij}\)] x_j\);
      \EndFor
    \EndFor
  \EndProcedure
\end{algorithm}

For the AFLP compression this approach already yielded the best performance. However, when using FPX a block-wise scheme
as in \cite{AmeJegExcMarPic:2025} showed slightly better results where (up to) 64 contiguous entries of a single column
of \(D\) are first decompressed into a local buffer.

The results when using Algorithm~\ref{alg:zdmvm} in \mcH-, \mcUH and \mcHH-matrix-vector multiplication are shown in
Figure~\ref{fig:zmvmtime}. There Algorithms~\ref{alg:phmvm}, \ref{alg:mvmuni} and \ref{alg:mvmh2} are used as they
showed the best performance in the experiments in Section~\ref{sec:mvm}. Results are again presented for an increasing
matrix size and a fixed accuracy of \(\varepsilon = 10^{-6}\) as well as for a fixed problem size and an increasing
accuracy. All results are given as speedups compared to the uncompressed \mcH, \mcUH and \mcHH-multiplication as shown
in Figure~\ref{fig:mvmtime}.

As can be seen, even with this simple implementation the performance improvement with compression is significant with a
speedup of about 2x up to 3x for \mcH-matrices. For \mcUH-matrices, this reduces to about 1.5x up to 2.5x and
is slightly less for \mcHH-matrices. In the latter case, actually no improvement was observed for the highest
accuracies. This picture reflects the difference in the reduction of the memory footprint due to compression, which is
best for \mcH-matrices and least for \mcHH-matrices. This also applies to the compression formats: AFLP shows
best performance even though the decompression costs are higher compared to the FPX scheme (see
Remark~\ref{rem:fpxspeed}). But since AFLP results in a better memory compression which leads to less bandwidth
utilization, the total performance gains are higher.

\begin{figure}[h]
  \centering
  \pgfdeclareimage[width=.43\textwidth]{rooflinezH}{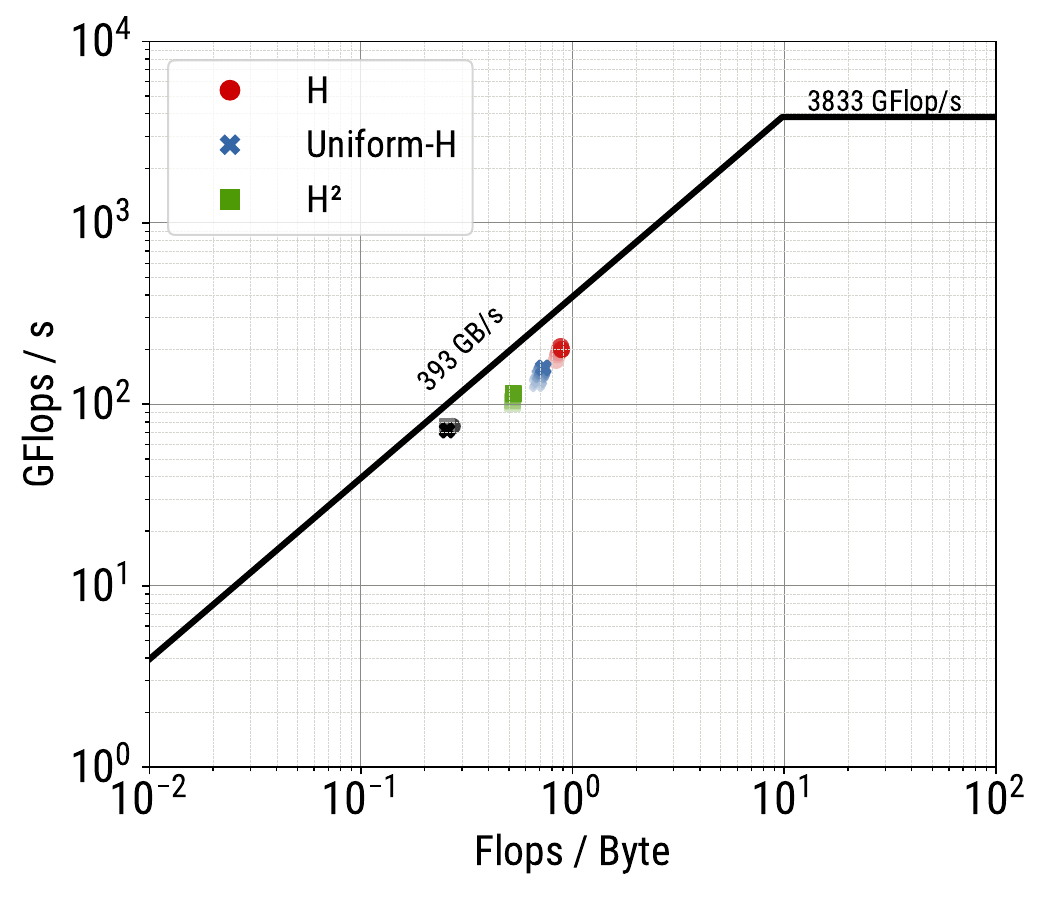}
  \pgfuseimage{rooflinezH}
  \caption{Roofline plot for \mcH-MVM, \mcUH-MVM and \mcHH-MVM using AFLP compression.}
  \label{fig:rooflinezH}
\end{figure}

\begin{figure}[h]
  \centering
  \pgfdeclareimage[width=.24\textwidth]{timevsdim}{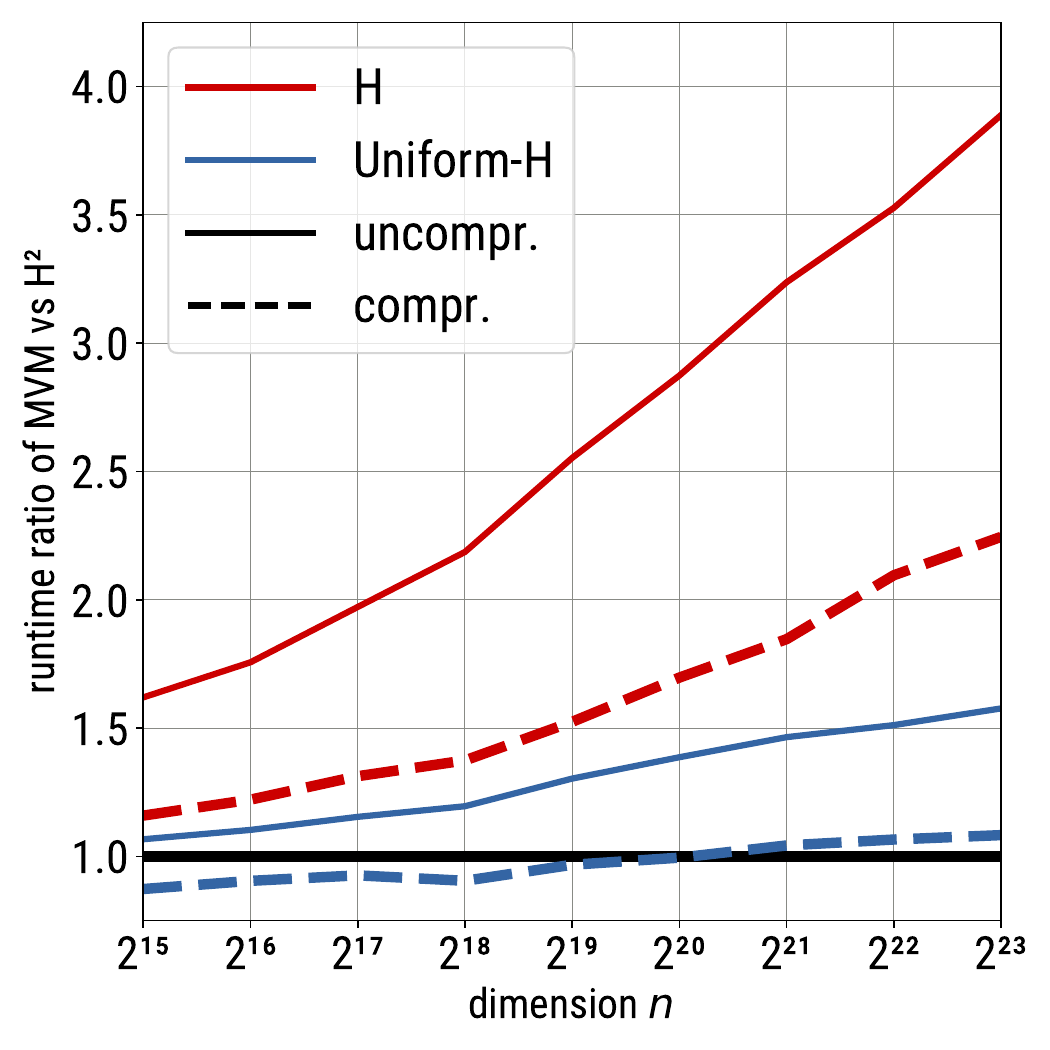}
  \pgfdeclareimage[width=.24\textwidth]{timevseps}{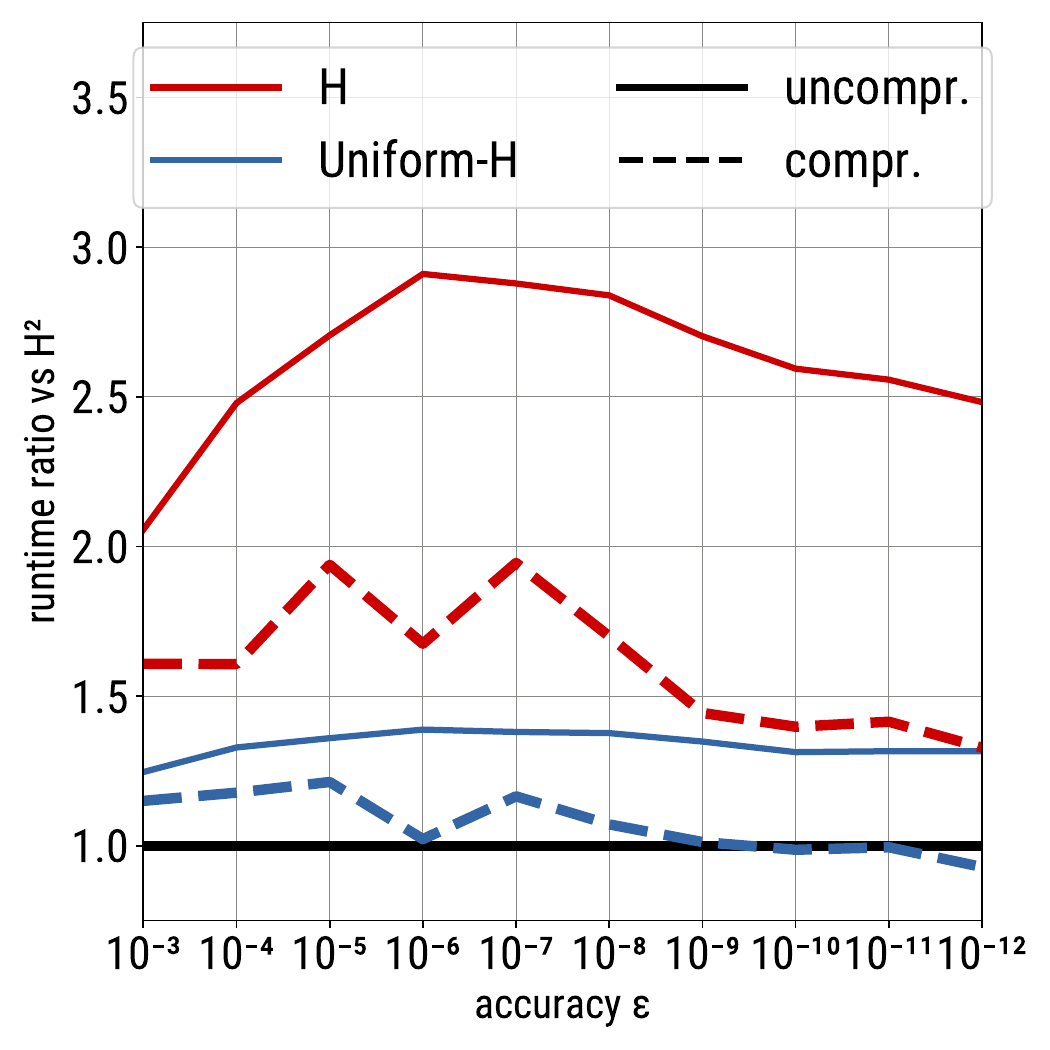}
  \pgfuseimage{timevsdim}\pgfuseimage{timevseps}
  \caption{Comparison of runtime of uncompressed and compressed (AFLP) \mcH- and \mcUH-matrix matrix-vector
    multiplication with \mcHH-format depending on matrix size (left) and accuracy (right).}
  \label{fig:timevsh2}
\end{figure}

For all matrix formats the advantage of compressed memory for matrix-vector multiplication reduces with an
increasing accuracy but increases with the problem size, with an exception for \mcHH-matrices. This again replicates the
compression rate results in Figure~\ref{fig:comprates}. However, in all cases the observed speedups are smaller compared
to the improvements due to compression, which is at least partly due to the decompression overhead. This overhead is
also visible in the roofline plot in Figure~\ref{fig:rooflinezH}. While the overall performance increases, there is a
wider gap to the optimal achievable performance than for uncompressed matrix-vector multiplication. Instead of the 80\%
of the peak performance of the uncompressed multiplication only about 60\% are reached with compression enabled.

As for the memory compression, we may further look into the advantage of the \mcHH-matrix format compared to \mcH- and
\mcUH-matrices for matrix-vector multiplication. Figure~\ref{fig:timevsh2} shows the ratio of the time for uncompressed
and compressed multiplication between \mcH-matrices/\mcUH-matrices and \mcHH-matrices, which represents the performance
penalty of the matrix formats compared to the \mcHH-matrix format. This is significantly reduced with compression
and brings \mcUH-matrices very close to \mcHH-matrices for the considered problem sizes and accuracies. However,
asymptotically the better runtime complexity of the \mcHH-matrix format will lead to the best performance.





\section{Conclusion} \label{sec:conclude}

Similar to \mcH-matrices, also the uniform \mcH-matrix format and the \mcHH-matrix format benefit significantly from an
optimized memory representation of the dense and low-rank data. However, this improvement is smaller compared to
\mcH-matrices since the data representation within this advanced formats is already optimized by shared and nested
cluster bases. Furthermore, when tightly coupling the compressed format with the matrix-vector multiplication all matrix
formats show a noticable performance speedup compared to the uncompressed case. Though not investigated in this work,
these runtime improvements should also apply to similar methods, e.g., triangular vector solves for \mcH-LU
factorizations. Even though a standard model problem was used in the this work, the results should not be limited to
this application, though the actual compression rates and thus the performance benefit for matrix-vector multiplication
depends on the distribution of singular values within low-rank blocks since these define the permitted accuracy for each
low-rank vector pair.



\printbibliography

\end{document}